\documentclass[draft]{agujournal2019}
\usepackage{url} %this package should fix any errors with URLs in refs.
\usepackage{lineno}
\usepackage[inline]{trackchanges} %for better track changes. finalnew option will compile document with changes incorporated.
\usepackage{soul}
\usepackage{physics}
\usepackage{accents}
\usepackage{xcolor}
\usepackage[normalem]{ulem} % for underlining with \uline (better line breaks than \underline)

\newcommand{\myhref}[2]{%
  \leavevmode
  \pdfstartlink user{
    /Subtype /Link /A <<
      /S /URI /URI (#1)
    >>
  }%
  \textcolor{blue}{\uline{#2}}%
  \pdfendlink
}

\newcommand{\bfu}{\boldsymbol{u}}

\newcommand{\bfF}{\boldsymbol{F}}

\newcommand{\nabp}{\nabla_{p}}

\newcommand{\ItI}{\textit{I}}

\draftfalse

\journalname{Geophysical Research Letters}

\begin{document}

%%%%%%%%%%%%%%%%%%%%%%%%%%%%%%%%%%%%%%%%%%%%%%%
%  TITLE
%
% (A title should be specific, informative, and brief. Use
% abbreviations only if they are defined in the abstract. Titles that
% start with general keywords then specific terms are optimized in
% searches)
%
%%%%%%%%%%%%%%%%%%%%%%%%%%%%%%%%%%%%%%%%%%%%%%%

% Example: \title{This is a test title}

\title{Understanding the Evolution of Global Atmospheric Rivers with a Vapor Kinetic Energy Framework}

%%%%%%%%%%%%%%%%%%%%%%%%%%%%%%%%%%%%%%%%%%%%%%%
%
%  AUTHORS AND AFFILIATIONS
%
%%%%%%%%%%%%%%%%%%%%%%%%%%%%%%%%%%%%%%%%%%%%%%%

% Authors are individuals who have significantly contributed to the
% research and preparation of the article. Group authors are allowed, if
% each author in the group is separately identified in an appendix.)

% List authors by first name or initial followed by last name and
% separated by commas. Use \affil{} to number affiliations, and
% \thanks{} for author notes.
% Additional author notes should be indicated with \thanks{} (for
% example, for current addresses).

% Example: \authors{A. B. Author\affil{1}\thanks{Current address, Antartica}, B. C. Author\affil{2,3}, and D. E.
% Author\affil{3,4}\thanks{Also funded by Monsanto.}}

\authors{Aidi Zhang\affil{1}, Da Yang\affil{1,2}, Hing Ong\affil{3}, Zhihong Tan\affil{4}}

% \affiliation{1}{First Affiliation}
% \affiliation{2}{Second Affiliation}
% \affiliation{3}{Third Affiliation}
% \affiliation{4}{Fourth Affiliation}

\affiliation{1}{Department of Geophysical Sciences, University of Chicago, Chicago, IL, USA}
\affiliation{2}{Department of Geophysics, Stanford University, Stanford, CA, USA}
\affiliation{3}{Independent Scholar, Davis, CA, USA}
\affiliation{4}{Atmospheric and Oceanic Sciences Program, Princeton University, Princeton, NJ, USA}
%(repeat as many times as is necessary)

% Corresponding author mailing address and e-mail address:

% (include name and email addresses of the corresponding author.  More
% than one corresponding author is allowed in this LaTeX file and for
% publication; but only one corresponding author is allowed in our
% editorial system.)

% Example: \correspondingauthor{First and Last Name}{email@address.edu}

\correspondingauthor{Da Yang}{dayang@stanford.edu}

\correspondingauthor{Aidi Zhang}{aidiz@uchicago.edu}

%%%%%%%%%%%%%%%%%%%%%%%%%%%%%%%%%%%%%%%%%%%%%%%
% KEY POINTS
%%%%%%%%%%%%%%%%%%%%%%%%%%%%%%%%%%%%%%%%%%%%%%%
%  List up to three key points (at least one is required)
%  Key Points summarize the main points and conclusions of the article
%  Each must be 140 characters or fewer with no special characters or punctuation and must be complete sentences

% Example:
% \begin{keypoints}
% \item	List up to three key points (at least one is required)
% \item	Key Points summarize the main points and conclusions of the article
% \item	Each must be 140 characters or fewer with no special characters or punctuation and must be complete sentences
% \end{keypoints}

\begin{keypoints}
\item We develop a vapor kinetic energy framework to examine atmospheric rivers' development across different ocean basins.

\item Potential energy (PE) to kinetic energy (KE) conversion is the primary driver sustaining atmospheric river intensity across all regions.

\item Eastward propagation is driven by vapor kinetic energy convergence, aided by PE-to-KE conversion near the North American west coast.

\end{keypoints}

%%%%%%%%%%%%%%%%%%%%%%%%%%%%%%%%%%%%%%%%%%%%%%%
%
%  ABSTRACT and PLAIN LANGUAGE SUMMARY
%
% A good Abstract will begin with a short description of the problem
% being addressed, briefly describe the new data or analyses, then
% briefly states the main conclusion(s) and how they are supported and
% uncertainties.

% The Plain Language Summary should be written for a broad audience,
% including journalists and the science-interested public, that will not have 
% a background in your field.
%
% A Plain Language Summary is required in GRL, JGR: Planets, JGR: Biogeosciences,
% JGR: Oceans, G-Cubed, Reviews of Geophysics, and JAMES.
% see http://sharingscience.agu.org/creating-plain-language-summary/)
%
%%%%%%%%%%%%%%%%%%%%%%%%%%%%%%%%%%%%%%%%%%%%%%%

%% \begin{abstract} starts the second page

\begin{abstract}
Atmospheric rivers (ARs) often cause damaging winds, rainfall, and floods. However, the physical mechanisms governing their evolution remain poorly understood. To close this gap, we perform a global Vapor Kinetic Energy (VKE) budget analysis. Using two formulations of VKE, we show that ARs are governed by similar mechanisms regardless of ocean basins. ARs intensify primarily through the conversion of potential energy to kinetic energy (PE-to-KE), with horizontal convergence of vapor kinetic energy providing a secondary contribution in some regions. ARs decay mainly through condensation and turbulent dissipation, while their propagation is governed by the downstream convergence and upstream divergence of vapor kinetic energy. We also find PE-to-KE conversion varies spatially and strengthens in regions of greater baroclinic instability or enhanced topographic lifting, e.g., along North America's west coast. Collectively, these findings demonstrate that the VKE framework provides a powerful diagnostic for how physical processes shape AR evolution and regional variability.
\end{abstract}

\section*{Plain Language Summary}
Atmospheric rivers (ARs) are narrow bands of fast-moving, concentrated water vapor in the mid-latitudes. They often bring strong winds, heavy rainfall, and flooding. In this study, we use an energy-based budget analysis to show that ARs follow a common pattern of evolution around the world. An AR grows when atmospheric instabilities convert potential energy into kinetic energy. It weakens when water vapor condenses into droplets and when turbulence dissipates its energy. Its movement is associated with the differences in energy flux between its upstream and downstream directions. We also find that the conversion of potential energy to kinetic energy depends on vertical air motion and its density anomaly, which is influenced by how unstable the atmosphere is and the shape of the land surface below.

%%%%%%%%%%%%%%%%%%%%%%%%%%%%%%%%%%%%%%%%%%%%%%%
%
%  BODY TEXT
%
%%%%%%%%%%%%%%%%%%%%%%%%%%%%%%%%%%%%%%%%%%%%%%%

\section{Introduction}\label{sec:intro}

Atmospheric rivers (ARs) are narrow streams of fast-moving, concentrated water vapor in the mid-latitudes. They account for the majority of meridional water vapor transport \cite{zhu1998proposed} and play a key role in precipitation events \cite{dettinger2011atmospheric,dettinger2013atmospheric,xiong2021influences}. Much like a coin with two sides, ARs can be both beneficial and hazardous. While they are essential for replenishing water supplies, they can also cause extreme precipitation, damaging winds, and other hazards (see \citeA{ralph2019scale} for a summary of AR-related hazards discussed in the literature).

Moving from this qualitative description of ARs and their effects toward a quantitative understanding requires defining a variable that reflects both their “moist” and “fast-moving” characteristics of AR. The most widely used variable is Integrated Vapor Transport (IVT; see Section~\ref{sec:AR_IVT} for details). Building on this, the Atmospheric River Tracking Method Intercomparison Project (ARTMIP) \cite{rutz2019atmospheric,lora2020consensus} has compared multiple AR detection algorithms, highlighting both areas of consensus and key differences. These analyses reveal that most algorithms agree on five regions with high AR frequency (North Pacific, Southeast Pacific, North Atlantic, South Atlantic, South Indian ocean, see figure~\ref{fig:1}). These regions form the focus basins for this study.

Leveraging these AR-detection algorithms, several statistical studies have demonstrated that ARs are associated with other atmospheric systems, such as extratropical cyclones and warm conveyor belts \cite{gimeno2014atmospheric,guo2020statistical}, extratropical anticyclones \cite{guo2020statistical}, Monsoon onset \cite{lee2019north,lee2021dynamics}, Rossby waves \cite{swenson2018resolution}, Rossby Wave Breaking \cite{ryoo2013impact,payne2014dynamics,mundhenk2016all,mundhenk2016modulation,hu2017linking}, annular modes\cite{baek2023atmospheric}, and extreme heat events \cite{scholz2024atmospheric}. 
Moreover, some research suggests that the AR frequency and its associated precipitation can be influenced by various large-scale climate patterns, including the Arctic Oscillation (AO) and the Pacific–North American (PNA) pattern \cite{guan20132010}, the North Atlantic Oscillation \cite{lavers2013nexus}, and the Madden–Julian Oscillation \cite{guan2012does}. Some of these studies have explicitly linked AR dynamics to these large-scale modes and synoptic systems. For example, \citeA{cordeira2013development} did a case study of two concurring AR that originates near a tropical cyclone and travel across the North-Pacific. They show that, during the early stage of AR development, the AR is associated with ascending moist air within an environment with high Convective Available Potential Energy and upper atmosphere divergence. While the ARs experience substantial precipitation losses as they travel across the North Pacific, they are replenished by ageostrophic circulation and IVT convergence. In addition, comparisons between ARs and local finite-amplitude wave activity (LWA) indicate that ARs are associated with the Rossby waves \cite{swenson2018resolution} and Rossby wave breaking (e.g., \citeA{hu2017linking} and \citeA{lee2021dynamics}.) Furthermore, \citeA{lee2021dynamics} demonstrate that a simple baroclinic instability model can successfully predict the locations of quasi-stationary ARs (QSAR). Recently, \citeA{lee2025quasi} show that the QSAR is a major source of Rossby wave in the subtropics.

However, a quantitative understanding of how different physical processes affect AR evolution remains limited. Recently, \citeA{ong2024vapor} introduced a budget analysis framework based on a new energy-focused AR variable, the Vapor Kinetic Energy (VKE).  Using a local composite analysis over the North Pacific, they demonstrated that AR growth is primarily driven by the potential energy (PE) to kinetic energy (KE) conversion, whereas AR decay is mainly due to turbulence dissipation and condensation processes. Their results also highlighted that AR movement is largely contributed by advection.

Although \citeA{ong2024vapor} provide a VKE budget analysis framework and identify key factors contributing to the growth, decay, and movement of ARs, their analysis is limited to a single AR composite at a limited region. Whether their findings are regionally dependent remains unclear. To investigate the universality and regional variation in the contributions by different AR tendencies to AR evolution, we perform a \textit{global} budget analysis across the five ocean basins with the highest AR frequency. We diagnose how each physical process contributes to AR growth, decay, and movement in these basins. We also discuss how the spatial variation of the main driver of AR growth, the PE-to-KE conversion, is influenced by baroclinic instability, baroclinic conversion, and topographic effects. 
In Section~\ref{sec:MandM}, we propose two different kinds of vapor kinetic energy and develop a budget analysis framework that can be used for diagnosing how each AR tendency term contributes to AR growth, decay, and movement during the AR life cycle. In Section~\ref{sec:results}, we present the result of the \textit{global} budget analysis. Section~\ref{sec:conclusion} summarizes our conclusions.

\section{Materials and Methods}\label{sec:MandM}

\subsection{Integrated Vapor Transport}\label{sec:AR_IVT}
Atmospheric rivers (ARs) are characterized by two essential components: fast-moving winds and abundant water vapor. Any quantitative study of ARs requires a variable that captures both of these ingredients. The most commonly used variable,  the Integrated Vapor Transport (IVT) \cite{gimeno2014atmospheric, payne2020responses,lora2020consensus}, is defined as,
\begin{eqnarray}
    \mathrm{IVT}\equiv\sqrt{\Big(-\frac{1}{g}\int_{p_B}^{p_T}qudp\Big)^2+\Big(-\frac{1}{g}\int_{p_B}^{p_T}qvdp\Big)^2} 
    \label{eq:IVT_def}
\end{eqnarray}
Where $q$ is the specific humidity; $u$ is the zonal velocity; $v$ is the meridional velocity; $g$ is the gravity constant; $p$ is the pressure; $p_B$ is the surface pressure; $p_T$ is the pressure level above which the moisture is negligible, which is set to be 200 hPa in this study.

Although IVT is widely used in the literature and its corresponding algorithms have been extensively explored in previous studies (e.g., \citeA{lora2020consensus}),  it is hard to derive its budget equation from equation~\eqref{eq:IVT_def}. This complexity makes it challenging to quantitatively analyze how individual processes affect AR evolution. 

\subsection{Vapor Kinetic Energy}\label{sec:AR_VKE}

\citeA{ong2024vapor} propose a vapor kinetic energy framework to study atmospheric rivers. In this study, we explore two different forms of vapor kinetic energy. To distinguish the forms, we refer to the variable proposed in \citeA{ong2024vapor} as the Vapor Transport Energy (VTE),
\begin{eqnarray}
    \mathrm{VTE} &\equiv& q^2(u^2+v^2)/2
    \label{eq:VTE_def}\\
    \mathrm{IVTE} &\equiv& -\frac{1}{g}\int_{p_B}^{p_T}q^2\frac{u^2+v^2}{2}dp
    \label{eq:IVTE_def}
\end{eqnarray}
For convenience, we define IVTE as the vertically integrated VTE.

The terminology VTE reflects its physical meaning. VTE is proportional to the square of the magnitude of the Vapor Transport ($\mathrm{VT}\equiv q\textbf{u}$).  If we consider the VT as a wave, then the VTE is the wave activity of the VT. The advantage of VTE is that it is governed by a straightforward budget equation, which allows a quantitative analysis of the roles different mechanisms play in AR evolution. \citeA{ong2024vapor} demonstrate the potential of this budget analysis through a composite study in the northwest Pacific and a case study in the northeast Pacific. 

Although the VTE is in the dimension of energy per unit mass (e.g., J/kg), it is not the kinetic energy carried by water vapor. Another natural choice of energy is the Kinetic Energy of Vapor (KEV), 
\begin{eqnarray}
    \mathrm{KEV} &\equiv& q(u^2+v^2)/2
    \label{eq:KEV_def}\\
    \mathrm{IKEV} &\equiv& -\frac{1}{g}\int_{p_B}^{p_T}q\frac{u^2+v^2}{2}dp
    \label{eq:IKEV_def}
\end{eqnarray}
Where IKEV is the vertically integrated KEV.

The definition of KEV has a clear physical meaning: it represents the kinetic energy carried by water vapor. In contrast, VTE is the wave activity associated with vapor transport. It shares many similarities with the VT because of the equal power of $q$, $u$, and $v$. As a result, VTE and VT are more sensitive to the moisture, and KEV is more sensitive to the KE. Both KEV and VTE allow for quantitative budget analyses using the following budget equations,
\begin{eqnarray}
    \pdv{\mathrm{VTE}}{t}&=&-\nabp\cdot(\bfu\cdot \mathrm{VTE})-\pdv{ }{p}\,(\omega\cdot \mathrm{VTE})-q^2\bfu\cdot\nabp\Phi+2qK\cdot S +q^2\bfu\cdot\bfF\label{eq:VTE_tend}\\
    \pdv{\mathrm{KEV}}{t}&=&-\nabp\cdot(\bfu\cdot \mathrm{KEV})-\pdv{ }{p}\,(\omega\cdot \mathrm{KEV})-q\bfu\cdot\nabp\Phi+K\cdot S +q\bfu\cdot\bfF\label{eq:KEV_tend}
\end{eqnarray}
Where $\bfu$ is the horizontal velocity, $\nabp$ is the horizontal nabla operator in the pressure coordinate, $\omega$ is the vertical velocity in the pressure coordinate, $\Phi$ is the geopotential, $K\equiv (u^2+v^2)/2$ is the KE per unit mass, $S$ is the source/sink of water vapor, $\bfF$ is the parameterized subgrid-scale forcing in the momentum equation. (Note that the subgrid-scale forcing varies among different models. In this study, the subgrid-scale forcing is from the tendency terms outputted from MERRA-2 reanalysis, see data section for details.)

The right-hand-side terms in equations~\eqref{eq:VTE_tend} and~\eqref{eq:KEV_tend} represent different physical processes that contribute to the evolution of VTE (and KEV, respectively). The first term corresponds to the horizontal convergence of the VTE (KEV) flux, while the second term represents the vertical convergence of the VTE (KEV) flux. The third term accounts for the contribution from ageostrophic motion, which also reflects the moisture-weighted conversion of potential energy (PE) to kinetic energy (KE). 
The fourth term represents KE-weighted sources and sinks of water vapor. Finally, the fifth term captures the effects of moisture-weighted subgrid-scale forcing in the momentum equation. 

\subsection{AR Composites and attributing to AR evolution}\label{sec:AR_composites}

In this study, we perform AR composite analyses at multiple locations along the axes where ARs most frequently occur across all five ocean basins. We use the method described in \citeA{ong2024vapor} to identify the main physical processes that contribute to the evolution of typical ARs at each location. We also examine how these processes differ among the ocean basins.
While the detailed procedure for finding the composite is in Text S2, we note that the composite analysis consists of a series of linear regressions on snapshots of the days when there is AR at the center of the composite domain. It only represents how a typical AR evolves in the composite domain during the regression period (from 2010 to 2019). While our method can be applied to subsets of ARs (i.e., ARs of different seasons), the composites of the AR subsets could evolve differently from the annual AR composite presented. We also acknowledge that the presented composite cannot show the evolution of extreme AR events, because the regression smears out the details of individual events.

To demonstrate how each tendency term contributes to AR evolution (we refer to AR evolution as the evolution of a typical AR, measured by a representative variable, from an Eulerian point of view in this study), we perform a composite analysis at a representative location in the North Pacific. More specifically, we center the composite domain ($150^\circ \times 30^\circ$) at ($179.7^\circ$W, $37.75^\circ$N), then construct a longitude–time (Hovmöller) diagram (Figure~\ref{fig:2}) by calculating the lead/lag composite on the AR intensity. 
The wide domain size is necessary for fully capturing the AR given its longitudinal drift over the 3-day lead/lag.
(Note that a larger domain can potentially include multiple ARs in some snapshots. However, the linear regression in our composite analysis filters out the parts of the fields that are not correlated with the AR centered in the domain. ARs elsewhere will affect the composite only if correlated with the AR at the center.) We then compute the latitude-averaged composites to produce the Hovmöller diagram.

To investigate regional variation (Figure~\ref{fig:3}), we repeat the composite analysis along the axes of most frequent AR occurrence in all five ocean basins \cite{lora2020consensus}. For each basin, we generate 50 composite sets by shifting the center of a fixed-size domain ($60^\circ \times 30^\circ$) along these straight lines (see Figure~\ref{fig:1}). For each domain position, we calculate a composite consisting of linear regressions of the vertically integrated KEV (VTE) tendency terms (equations~\eqref{eq:VTE_tend} and~\eqref{eq:KEV_tend}) against the mean IKEV (IVTE) value within a $1^\circ$ box around the domain center. The calculation procedure is identical for all domains (see Text S2).

To quantitatively assess how the composite of a tendency term $X$ influences the evolution of the AR variable $I$, we follow a methodology used in several regression studies \cite{lim1991structure,chang1993downstream,andersen2012moist,ong2024vapor}. The contribution from $X$ to the growth or decay of $I$ is calculated as follows,
\begin{eqnarray}  
\mathrm{Contribution\,to\,growth/decay}\equiv\iint_S X\cdot I dS\,\Big{/}\iint_S I^2 dS\label{eq:GD_def}
\end{eqnarray}
Where $\iint_S dS$ is the area integral in the composite domain $S$. Note that both $I$ and $X$ are already vertically integrated before doing the composite analysis. With respect to the AR movement, the contribution is defined as, 
\begin{eqnarray}    \mathrm{Contribution\,to\,movement}&\equiv&\iint_S X\cdot\tilde{I}_t dS\,\Big{/}\iint_S \tilde{I}_t^2 dS\label{eq:MM_def}\\
    \tilde{I}_t&\equiv& \pdv{I}{t} - \frac{\iint_S (\ItI\cdot\partial I/\partial t) dS}{\iint_S I^2 dS}I\label{eq:MM_def_2}   
\end{eqnarray}
We define $\tilde{I}_t$ in equation~\eqref{eq:MM_def_2} using Gram–Schmidt normalization to remove the net growth or decay from $\partial I/\partial t$. Physically speaking, subtracting the second term on the right-hand side of equation~\eqref{eq:MM_def_2} is equivalent to removing the uniform-rate AR growth/decay. The resulting tendency $\tilde{I}_t$ is orthogonal to $I$ and represents the AR movement. We then project $X$ onto $\tilde{I}_t$, as shown in equation~\eqref{eq:MM_def}, to quantify its contribution to AR movement. With equation~\eqref{eq:MM_def}, we can quantify how each tendency term contributes to AR movement throughout its life cycle, even when the AR is growing/decaying. Note that a positive contribution to movement also indicates that $X$ increases $I$ more in the downstream direction of the AR than in the upstream direction, representing effective AR movement. (Throughout this paper, “upstream” and “downstream” are defined relative to the overall AR displacement, rather than the local vapor velocity.)

\subsection{Data}\label{sec;data}

In this study, we use reanalysis data from the Modern-Era Retrospective analysis for Research and Applications, Version 2 (MERRA-2). Specifically, we utilize four sets of 3-hourly 3D MERRA-2 data: the instantaneous pressure-level assimilated meteorological fields; the time-averaged model-level assimilated meteorological fields; the time-averaged pressure-level assimilation moist tendencies; and the time-averaged pressure-level assimilation wind tendencies \cite{GMAO2015inst,GMAO2015tavg,GMAO2015qdt,GMAO2015udt}. Model-level data are interpolated to pressure levels when calculating the tendency terms and composites. For consistency with \citeA{ong2024vapor}, we use data from 2010 to 2019. Note that MERRA-2 reports moisture sources and sinks from three key components: moist processes, turbulence, and chemistry. Similarly, the parameterized KE tendency is reported from turbulence, gravity wave drag, and convection. (See \citeA{Bosilovich2016} for details.) Accordingly, we decompose $S$ and $\bfF$ in equations~\eqref{eq:VTE_tend} and~\eqref{eq:KEV_tend} based on these components. 

Although this study relies solely on MERRA-2 reanalysis data, \citeA{ong2024vapor} have conducted budget analyses using both MERRA-2 and ERA-5 data and have found similar results in both cases. Therefore, we do not expect a major difference with different sets of reanalysis data.

Distinguishing ARs from tropical cyclones (TCs) solely by AR-related variables is challenging due to overlapping moisture and dynamical features. To minimize the influence of tropical and subtropical cyclones, we utilize TC data from the International Best Track Archive for Climate Stewardship (IBTrACS) project \cite{knapp2010international}. We exclude systems that are marked as tropical cyclones, tropical storms, tropical depressions, or subtropical cyclones in IBTrACS, thereby ensuring that their effects are removed from our AR analysis. Fields within a $4^\circ$ circle around the systems' center are considered non-AR systems and are excluded from composite calculations. Figure S1 demonstrates that this method is sufficient to remove the signal of tropical cyclones.

\section{Results}\label{sec:results} 

\subsection{AR Detectability with Different Variables}\label{sec:AR_detectability}

Before conducting the AR budget analysis using IKEV and IVTE, we first examine whether the AR detected by these variables is consistent with the existing detection algorithms in the literature. The similarity between IVTE and IVT has been examined in \citeA{ong2024vapor}. They developed IVTE-based AR detection algorithms using two different ARTMIP members \cite{mundhenk2016all,ullrich2021tempestextremes} and show that the IVTE-based AR detection algorithms yield similar results as the IVT-based algorithms. In this study, we further develop the TempestExtremes-based algorithm \cite{ullrich2021tempestextremes, ong2024vapor} and apply it to both the IKEV and IVTE. The threshold of each variable is chosen to get a similar mean AR frequency globally. The details of our detection algorithm are described in Text S1. 

\begin{figure}[h!]
    \centering
    \includegraphics[width=1\linewidth]{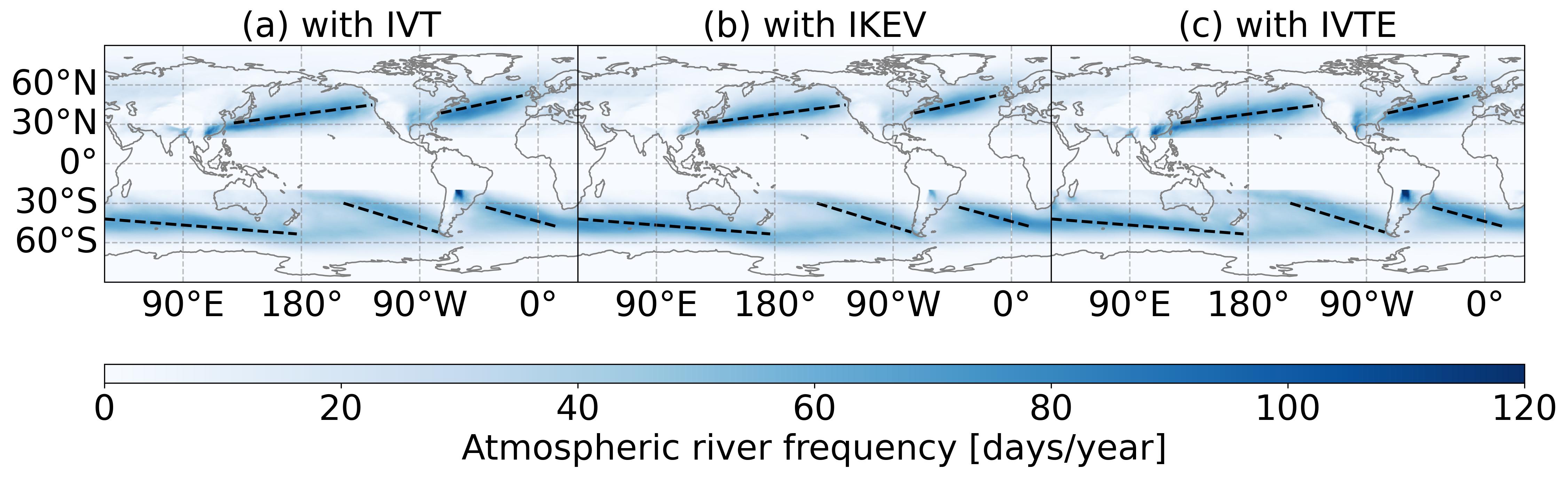}
    \caption{AR frequency with different variables in 2010-2019. Panel a shows the AR frequency with IVT. Panel b shows the AR frequency with IKEV. Panel c shows the AR frequency with IVTE. The tropical region between $\pm20^\circ$ latitude is not considered as AR in the detection. The black dashed lines are the representative locations of the five frequent-AR ocean basins.}
    \label{fig:1}
\end{figure}

To demonstrate that the vapor kinetic energy
can detect ARs as effectively as the IVT, figure~\ref{fig:1} shows the mean atmospheric river (AR) frequency from different variables. Panel (a) shows the one with IVT; panel (b) shows the one with IKEV; and panel (c) shows the one with IVTE. The AR frequency maps are visually similar across all three variables. Our algorithm consistently identifies the five consensus regions of frequent AR activity (the North Pacific, Southeast Pacific, Sorth Atlantic, South Atlantic, and South Indian Ocean) as described in \citeA{lora2020consensus}. The black dashed lines are the representative locations of the five frequent-AR ocean basins, which are used in the budget analysis in section~\ref{sec:AR_ana}.

In Figures S2 and S3, we provide an additional comparison of ARs detected using the three different variables. Comparing the annual frequency (Figure S2a) and seasonal frequency (Figure S3), the ARs detected via IKEV capture fewer AR features in lower latitudes in both Northern and Southern Hemispheres and more AR features in higher latitudes in Southern Hemisphere.
This is natural because IVT/IVTE is more sensitive to moisture and IKEV is more sensitive to wind. The differences arise from the strong equator-to-pole moisture gradient and the stronger jet stream in the Southern Hemisphere. The difference also provide an alternative approach to distinguish the windy part of AR from the moist part of AR. Even though the AR-frequency difference among different AR variables is not small, the difference is comparable to differences between ARTMIP members. In Figure S2bc, we plot two snapshots studied in Figure 4 of \citeA{lora2020consensus}, the differences between variables are not larger than the differences among the various IVT-based algorithms in ARTMIP.

\subsection{The Budget Analysis of AR Evolution}\label{sec:AR_ana}

\begin{figure}[h!]
    \centering
    \includegraphics[width=1\linewidth]{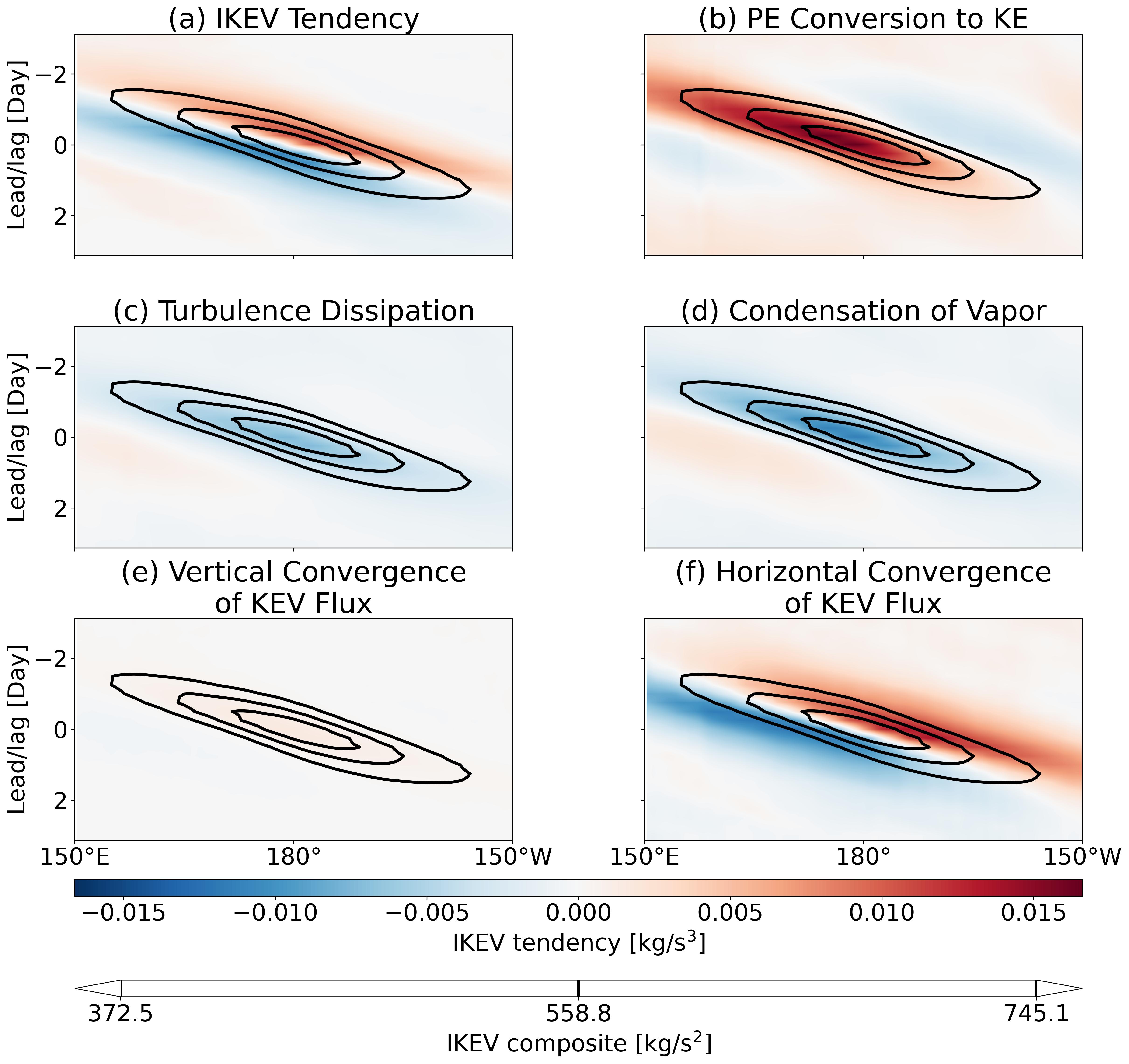}
    \caption{A Hovmöller diagram on the latitudinal-averaged (from 22.75$^\circ$N to 52.75$^\circ$N) AR composite in North Pacific. The black contours show the IKEV composite. The colored fields indicate the regressed vertically-integrated IKEV tendency variable. Panel (a) shows IKEV tendency; panel (b) shows the PE conversion to KE;  panel (c) shows the turbulence dissipation; panel (d) shows the condensation of vapor; panel (e) shows the vertical convergence of the KEV flux; panel (f) shows horizontal convergence of the KEV flux. 
    \label{fig:2}
    }
\end{figure}

To show how ARs evolve in time (in an Eulerian perspective, see Section~\ref{sec:AR_composites}), Figure~\ref{fig:2} shows the longitude–time (Hovmöller) diagram in the North Pacific. In all panels, the black contours represent the IKEV composite. Its eastward tilt reflects the eastward movement of the AR. This means that, in a latitudinal mean sense,  the eastern side corresponds to the downstream region of the AR, while the western side corresponds to the upstream region. The colored fields in each panel are the tendency terms. The Hovmöller diagrams in figure~\ref{fig:2} largely replicate the findings from the zero-lag composites in \citeA{ong2024vapor}. More specifically, AR growth is primarily driven by the PE-to-KE conversion (Figure~\ref{fig:2}b). Its dissipation mainly comes from turbulence dissipation (Figure~\ref{fig:2}c) and vapor condensation (Figure~\ref{fig:2}d). These findings holds as AR evolve in the non-zero lead/lag.

What is slightly different from \citeA{ong2024vapor} is that we consider the flux form in equations \eqref{eq:VTE_tend}-\eqref{eq:KEV_tend}, whereas \citeA{ong2024vapor} use advection form. Here, the integrated vertical convergence of KEV flux (Figure~\ref{fig:2}e) represents the net vertical flux ($\omega\cdot\mathrm{KEV}$) from near the surface and the top of domain. The KEV flux near the surface is small because of the weak vertical velocity. The KEV flux at the domain top is small because of the negligible moisture. Hence, there is negligible net vertical flux of KEV. Similarly, the integrated horizontal convergence of KEV flux (Figure~\ref{fig:2}f) combines the effect of both the horizontal advection and vertical advection. In \citeA{ong2024vapor}, the former is the primary contributor to the AR movement and the latter is the secondary contributor to AR growth. Both of them can be seen in Figure~\ref{fig:2}f. (Note that the vertical convergence of KEV flux can redistribute KEV from lower to upper troposphere despite being negligible in an integrated sense. See figure~S10cf as an example.)

The findings presented in Figure~\ref{fig:2} are consistent with those of \citeA{ong2024vapor}. However, both analyses rely on a single, hand-picked composite domain in the North Pacific. To assess the generality of these results, we perform a \textit{global} composite analysis using representative locations across the five major AR regions, as shown in Figure~\ref{fig:1}. We conduct budget analyses using both IVTE and IKEV, finding that the results are very similar. The similarity suggests that the partition of $q$ does not change the presented findings. For simplicity, all results of the budget analysis presented in the main text are based on the IKEV, while the IVTE-based analysis is provided in the supplemented material.

Figure~\ref{fig:3} shows the contributions from IKEV tendency terms to AR growth/decay (panel~a) and to AR movement (panel~b). Each vertical panel depicts the contributions from the major tendency terms in each basin, with horizontal arrows marking the center of each composite. Overall, the AR budget composition is similar across all basins.

\begin{figure}[h!]
    \centering
    \includegraphics[width=1\linewidth]{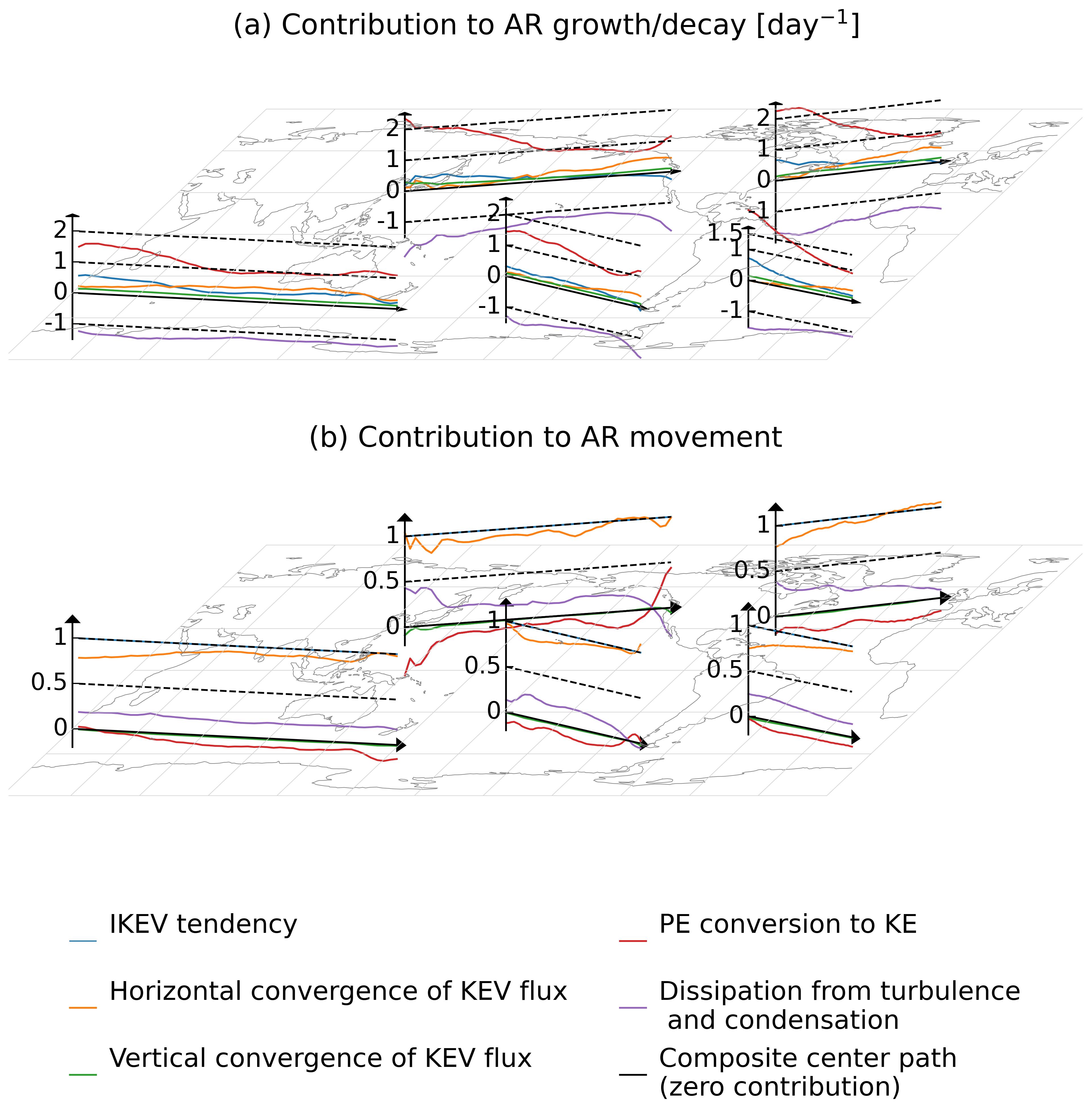}
    \caption{The global IKEV budget analysis on the AR evolution. Panel (a) shows the contribution to the growth/decay. Panel (b) shows the contribution to the movement. Each vertical panel shows the contribution from the tendency terms in each basin. The horizontal arrows indicate the center of each composite. The red is for the PE conversion to KE. The orange is for the horizontal convergence of the KEV flux. The green is for the vertical convergence of KEV flux. The purple is for the dissipation from turbulence and condensation. The blue is for the net IKEV tendency. (A multi-paneled version is provided in Figure S6)}
    \label{fig:3}
\end{figure}

Starting with Figure~\ref{fig:3}a and using the North Pacific as an example, the PE-to-KE conversion (red) is the dominant growth source. Its magnitude decreases from west to east across the basin, but increases slightly again near the North American continent. This west-to-east decreasing trend appears in all basins, while the near-coast increase on the eastern side occurs only where a major continent is present (see also South America and the southeast Pacific). The horizontal convergence of KEV flux (orange) in the western Pacific is negligible, but increases eastward, becoming a secondary growth source. This eastward increase is consistent across other basins except in the Indian Ocean, where it peaks in the mid-basin and is smaller at the basin edges. The vertical convergence of KEV flux (green) remains negligible in all basins.
 
The dissipation from turbulence and condensation (purple) is the major decay term. In the North Pacific, it is particularly large and negative in the west, increasing toward the east but decreasing again near the North American coast. The west-to-east increase occurs in all basins, but the eastern decrease is apparent only in the northeast and southeast Pacific. Notably, this trend is the opposite of that for PE-to-KE conversion, indicating that part of the KEV generated from PE-to-KE conversion is locally dissipated by turbulence and condensation.

In Figure~\ref{fig:3}b, the horizontal convergence of KEV flux (orange) dominates AR movement in all basins. This reflects the downstream convergence and upstream divergence patterns shown in Figure~\ref{fig:2}f. The vertical convergence of KEV flux (green) is negligible for AR movement. 
In the northwest Pacific, the PE-to-KE conversion (red) contributes negatively, while dissipation (purple) contributes positively. The signs of these terms reflect a downstream–upstream asymmetry (see equation~\eqref{eq:MM_def}). In most basins, stronger PE-to-KE conversion on the west side produces a negative contribution to movement, with the northeast and southeast Pacific as exceptions, where stronger PE-to-KE conversion on the east side gives a positive contribution to the movement. The west–east trend of dissipation (purple) is opposite to that of PE-to-KE conversion, consistent with the relationship seen in panel~(a).

\subsection{The Regional Variation of the PE Conversion to KE} \label{sec:PE2KE}

Here, we want to answer the following questions: 
\begin{enumerate}
    \item Why does the PE-to-KE conversion typically contribute more to AR growth on the west side of the basins? (Q1)
    \item Why does the PE-to-KE conversion contribute more to AR growth and eastward movement as ARs approach North America from the Pacific? (Q2)
\end{enumerate}

To address Q1, we compare the spatial variation of PE-to-KE conversion with the Eady growth rate (EGR, see section 9.5 of \citeA{vallis2017atmospheric}), a classical, dry metric of baroclinic instability (See Text S4, Figure S11-S12). We find a pronounced west–east gradient in EGR. In regions of higher EGR, the baroclinic instability is expected to be stronger, leading to stronger PE-to-KE conversion and its contribution to the AR growth. Note that our discussion is limited to dry baroclinic instability. Previous studies \cite{emanuel1987baroclinic,lambaerts2012moist} have shown that moisture increases the baroclinic instability growth rate through latent heat release during condensation. However, defining a moist counterpart to the EGR remains challenging. 

\begin{figure}[htbp!]
    \centering
    \includegraphics[width=1\linewidth]{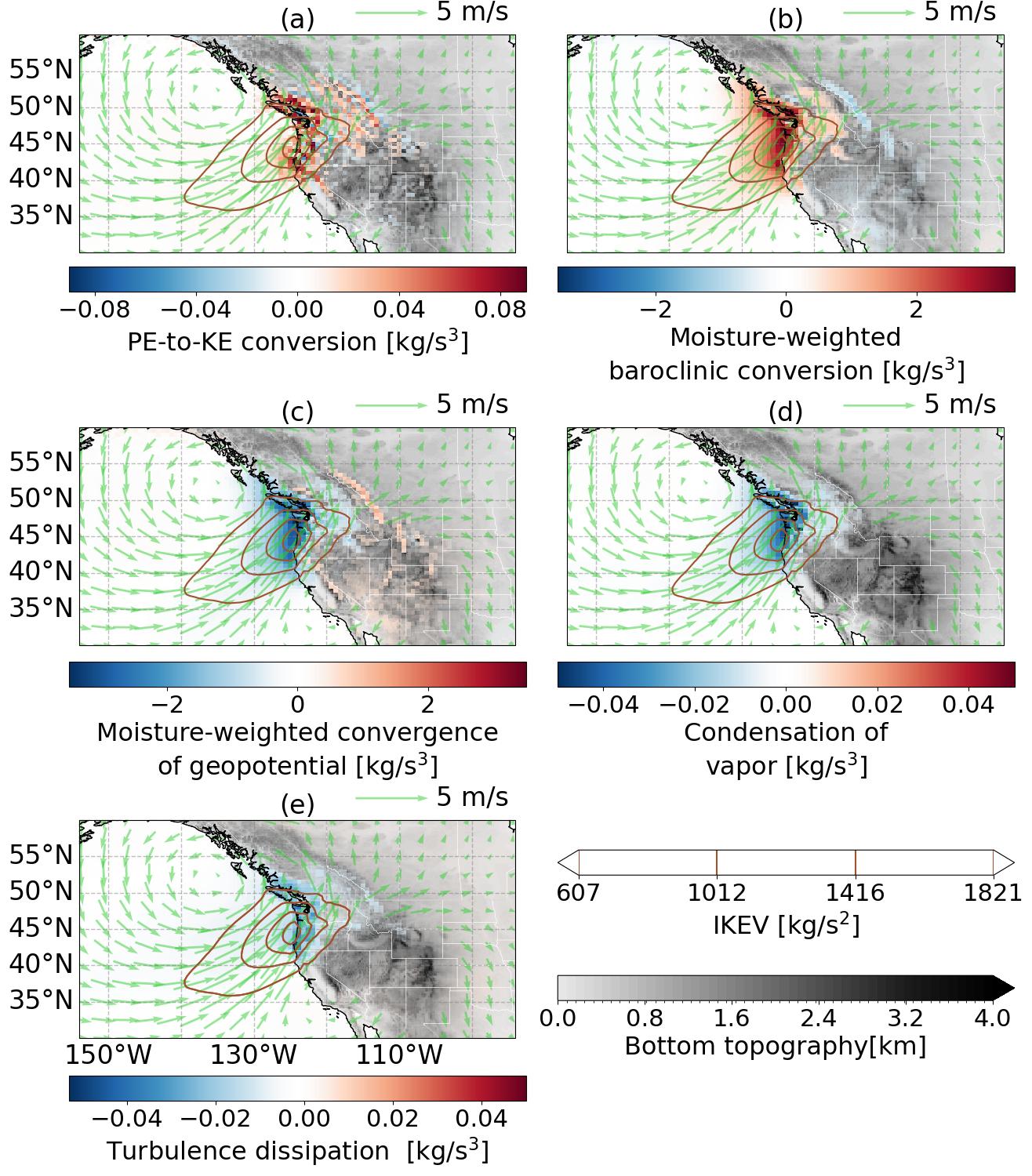}
    \caption{The composites of IKEV tendencies in the northeast Pacific. In all panels, the green arrows are the $850$~hPa wind profile and the grey shadings are the height of the topography. The brown contours are the IKEV composites (The levels are at 30\%, 50\%, 70\%, and 90\% of the maximum IKEV values). The colored fields correspond to different composites. Panel (a) shows the vertically-integrated PE-to-KE conversion. Panel (b) shows the vertically-integrated MWBC. Panel (c) shows the vertically-integrated moisture-weighted convergence of geopotential. Panel (d) shows the vertically-integrated condensation of vapor. 
    Panel (e) shows the vertically-integrated turbulence dissipation. Note that these tendencies are moisture-weighted, and the color ranges of each panel are different.
    }
    \label{fig:4}
\end{figure}

To address Q2, we calculate the composites of IKEV tendencies in the northeast Pacific (Figure~\ref{fig:4}). The AR is associated with a cyclonic flow to its northwest and an anticyclonic flow to its southeast, consistent with the pattern discussed in \citeA{guo2020statistical}. There is a cold front on the southeast side of the cyclonic flow, which roughly coincides with the AR composite.

In figure~\ref{fig:4}a, the colored field is the composite of the PE-to-KE conversion. It is much larger on the eastern side within the AR and on its northeastern (downstream) side. This stronger PE-to-KE conversion on the downstream side within the AR leads to a positive contribution to AR movement, as shown in the northeast Pacific in Figure~\ref{fig:3}b. Stripes of positive and negative regions also appear northeast of the AR. These stripes slant from northwest to southeast, following the regional topography, suggesting a significant role of topography in elevating PE-to-KE conversion near North America in Figure~\ref{fig:3}a,b. To further investigate the underlying mechanism, we decompose the PE-to-KE conversion terms into flux form. Specifically, we decompose the PE-to-KE conversion term in equation~\ref{eq:KEV_tend} by 
\begin{eqnarray}
    -q\bfu\cdot\nabp\Phi = -q\omega\alpha-q\nabla\cdot(\bfu_3\Phi)\label{eq:geop_flux_form}
\end{eqnarray}
Where $\alpha$ is the specific volume, $\bfu_3$ is the velocity in 3D, and $\nabla$ is the 3D nabla operator. On the right-hand side of the equation, the first term represents the moisture-weighted baroclinic conversion (MWBC) \cite{lorenz1955available,lorenz1967nature}. The second term represents the moisture-weighted convergence of geopotential, which includes both the horizontal and vertical convergence.

Figure~\ref{fig:4}b shows the composite of the MWBC, which reflects a triple correlation between $q$, $\omega$, and $\alpha$. Within the AR, the MWBC is positive (brown contours), consistent with \citeA{ong2024vapor}. Since moisture is higher within the AR, a stronger MWBC also implies stronger baroclinic conversion overall. This anomalously strong baroclinic conversion further indicates rising air is lighter (both higher temperature and moisture content can contribute to the lightness \cite{yang2022substantial,yang2023vapor,seidel2024vapor}). Fundamentally, this is associated with extratropical cyclones and accompanying front systems. For example, the front would allow cold air to intrude and force warm air up. This effect might be further amplified by latent heat release during condensation of the rising moist air.  Northeast of the AR, baroclinic conversion is positive on the windward side of the mountains and negative on the leeward side; these features match the spatial scale of topographic variation in the region. 

Figure~\ref{fig:4}c shows that much of the MWBC is compensated by the moisture-weighed convergence of geopotential. The cancellation can also be quantified by their contributions to AR growth/decay and movement. Using the same unit as figure~\ref{fig:3}, the MWBC results in $\sim74~\mathrm{day^{-1}}$ of AR growth and $6.6$ of AR movement, whereas the moisture-weighed convergence of geopotential contributes to $\sim-72~\mathrm{day^{-1}}$ of AR growth and $-6.9$ of AR movement. For comparison, the PE-to-KE conversion results in $\sim1.1~\mathrm{day^{-1}}$ of AR growth and $0.4$ of AR movement. Our qualitative interpretation of the processes governing AR movement is robust. But some quantitative discrepancy remains, likely due to the rapid changes in bottom topography in this region. (The residual appears to be associated with the large terms on the right hand side of equation~\eqref{eq:geop_flux_form}. It does not significantly affect our confidence in calculating the PE conversion to KE or in evaluating the related budget equations \eqref{eq:VTE_tend}-\eqref{eq:KEV_tend}, where the PE–KE conversion was explicitly calculated. The overall residual constitutes only a few percent of the PE–KE term.) 
Note that the MWBC is stronger than the moisture-weighed convergence of geopotential because of the energy converts from the available potential energy (APE) to KE in the Lorenz cycle \cite{lorenz1955available}. And the PE-to-KE conversion reflects the net effect of the baroclinic energy conversion, which energizes the AR.

Additionally, the stronger PE-to-KE conversion near the west coast of North America does not necessarily mean that the AR is stronger. In fact, the dissipation also becomes stronger in the region: Figure~\ref{fig:4}d shows enhanced condensation toward the continent, likely associated with topographic lifting and precipitation, and Figure~\ref{fig:4}e shows that the turbulent dissipation also becomes stronger on the windward side of the mountain. As a result, the PE-to-KE conversion is overcompensated by dissipation from the condensation and the turbulence, leading to the decay of ARs in this region (See the blue curve in the northeastern Pacific of figure~\ref{fig:3}a.)

\section{Conclusions}\label{sec:conclusion}

In this study, we use an energy framework to diagnose how different physical processes affect AR evolution and their regional dependence. We perform a \textit{global} budget analysis of ARs using two different versions of vapor kinetic energy: the Integrated Kinetic Energy of Vapor (IKEV) and the Integrated Vapor Transport Energy (IVTE). Our results reveal that AR evolution follows a broadly consistent pattern worldwide. Some of our findings were also found in some regional-specific studies on AR dynamics.

Globally, we find that the AR growth is primarily driven by the conversion of potential energy (PE) to kinetic energy (KE), while decay is dominated by energy losses from turbulence and condensation processes. AR movement is associated with downstream convergence and upstream divergence of KEV (or VTE) fluxes. The findings are qualitatively consistent with the evolution schematic in a case study about two TC-related ARs in the North Pacific (See figure 11 in \citeA{cordeira2013development}). They find that in the central North Pacific, the AR is associated with a strong thermally direct ageostrophic circulation (our PE-to-KE conversion) and IVT convergence (our IVKE convergence), which together offset the deleterious effect of high precipitation rate. This picture is consistent with our findings in the North Pacific Panels in Figure 3. However, the present study and \citeA{ong2024vapor} further quantify the contribution of each physical processes to the AR evolution with the IVKE framework.

Furthermore, we demonstrate that spatial variations of PE-to-KE conversion, the main driver of the AR growth, closely relate to other physical processes. Notably, the PE-to-KE conversion is modulated by the topography, which leads to a stronger PE-to-KE conversion near the west coast of North America (See figure 4). We also show that the PE-to-KE conversion is the aftermath of cancellation of positive baroclinic conversion and negative convergence of geopotential. Note that this pattern was also found in the central North Pacific \cite{ong2024vapor} and Northwestern Pacific (the initial AR-development stage in \citeA{cordeira2013development}). In addition, we find that the PE-to-KE conversion is stronger in the regions of higher Eady growth rate (more baroclinic unstable, see figures S11 and S12). \citeA{lee2021dynamics} show that the baroclinic instability is critical in the formation of quasi-stationary AR and can predict its location. Our findings indicates that it is also important for ARs in general.
 
Studying ARs from the lens of the vapor kinetic energy (VKE) aims to provide a firmer footing in atmospheric dynamics. However, important questions remain. For example, previous studies have shown that AR dynamics is closely related to Rossby wave \cite{swenson2018resolution} and Rossby wave breaking \cite{hu2017linking, lee2021dynamics}. This study is AR-centric and it is unclear how our results relates to Rossby wave breaking metrics such as the local finite-amplitude wave activity (LWA). Furthermore, it is unknown how ARs' VKE budget might change in a warming climate and help explain ARs' characteristics in the future. Will the relative contributions of individual physical processes in the VKE budget systematically shift as the climate warms? What roles will different meteorological systems (e.g., extratropical cyclones and frontal systems) and related dynamical processes play in these changes? And what controls the strength, frequency, or number of ARs? Addressing these open questions will require targeted experiments using a model hierarchy, ranging from idealized models to fully coupled general circulation models (GCMs), which we plan to pursue in future work.

\newpage
\section*{Open Research Section}
The MERRA2 data used in this study are publicly available \cite{GMAO2015inst,GMAO2015tavg,GMAO2015udt, GMAO2015qdt}. The tropical cyclone data is from IBTrACS \cite{gahtan2024ibtracs}. The dataset can be accessed via \myhref{https://zenodo.org/records/17290141?preview=1&token=eyJhbGciOiJIUzUxMiJ9.eyJpZCI6ImZjOWI3N2IwLThiNTgtNDFlOC1hOWUwLWExMDc2MjI2YzlmMCIsImRhdGEiOnt9LCJyYW5kb20iOiJjZDY3ZTBhZjA5NDMwMGUwODUwM2JlYmIxYjM2ZjgwYSJ9.plb2rkb8CpzQFeETX6NGiRFmDKLchG2JTOVOKKQDkh4x82WqDVkE9H0dl7quTxFCnIOTFJSRGvOCBY5dO2KShQ}{link}. We also provide the full text of the link in file \verb|link_to_dataset.txt| of this submission. The dataset and code used to generate the figures in this study will be made publicly available after the manuscript is accepted.

\section*{Supporting References}
The following references are cited in the Supporting Information for this article: 
\citeA{ong2024vapor,rutz2019atmospheric,ralph2019artmip,knapp2010international,gahtan2024ibtracs,lora2020consensus,zhu1998proposed,payne2020responses,lee2021dynamics,vallis2017atmospheric,simmonds2009biases,simmonds2021trends,gillett2021tropical}.

% This section MUST contain a statement that describes where the data supporting the conclusions can be obtained. Data cannot be listed as ''Available from authors'' or stored solely in supporting information. Citations to archived data should be included in your reference list. Wiley will publish it as a separate section on the paper’s page. Examples and complete information are here:
% https://www.agu.org/Publish with AGU/Publish/Author Resources/Data for Authors

\section*{Conflict of Interest Disclosure}
The authors declare there are no conflicts of interest for this manuscript.

\acknowledgments

This work was supported by a Packard Fellowship in Science and Engineering and an NSF CAREER Award (AGS-2048268) to Da Yang. It is also supported by the computational resources from the National Center for Atmospheric Research (NCAR) under project UCDV0026. We appreciate the helpful discussion with Noboru Nakamura and Malte Jansen. We also appreciate the comments from two anonymous reviewers.

Zhihong Tan is supported by awards NA18OAR4320123 and NA23OAR4320198 from the National Oceanic and Atmospheric Administration, U.S. Department of Commerce. The statements, findings, conclusions, and recommendations are those of the authors and do not necessarily reflect the views of the National Oceanic and Atmospheric Administration, or the U.S. Department of Commerce.

\bibliography{AR.bib}

\newpage
\renewcommand{\thefigure}{S\arabic{figure}}
\setcounter{figure}{0}
\title{Supporting Information for "Understanding the Evolution of Global Atmospheric Rivers with Vapor Kinetic Energy Framework"}

\noindent\noindent\textbf{Contents of this file}
%%%Remove or add items as needed%%%
\begin{enumerate}
\item Text S1 to S4
\item Figures S1 to S12
% \item Tables S1 to Sx
%if Tables are larger than 1 page, upload as separate excel file
\end{enumerate}

\noindent\textbf{Text S1. The AR detection algorithms}

The AR detection algorithm used in this study has two major differences compared with the one in \citeA{ong2024vapor}: the AR thresholds and the treatment of tropical cyclones (TCs) and similar moist systems.

The AR detection algorithm in \citeA{ong2024vapor} used a fixed maximum-Laplacian threshold ($\laplacian \mathrm{IVT} < -40,000\,\mathrm{kg\,m^{-1}s^{-1}rad^{-2}}$; $\laplacian \mathrm{IVTE} < -2,000\,\mathrm{kg\,s^{-2}rad^{-2}}$). To make the threshold more flexible for climate variation studies, we replace these fixed numerical thresholds with percentile thresholds. To be consistent with \citeA{ong2024vapor}, a flow is considered a potential AR candidate if its $\laplacian \mathrm{IVT}$ falls within the lowest 27th percentile. For IVTE (or IKEV) detections, the flow qualifies as a potential AR candidate if its Laplacian is within the lowest 20th percentile. In addition to the Laplacian criterion, we require the AR area to encompass at least 50 connected grid points and exclude points located between $20^\circ$N and $20^\circ$S. These additional conditions are consistent with those applied in \citeA{ong2024vapor}.

\begin{figure}
    \centering
    \includegraphics[width=1\linewidth]{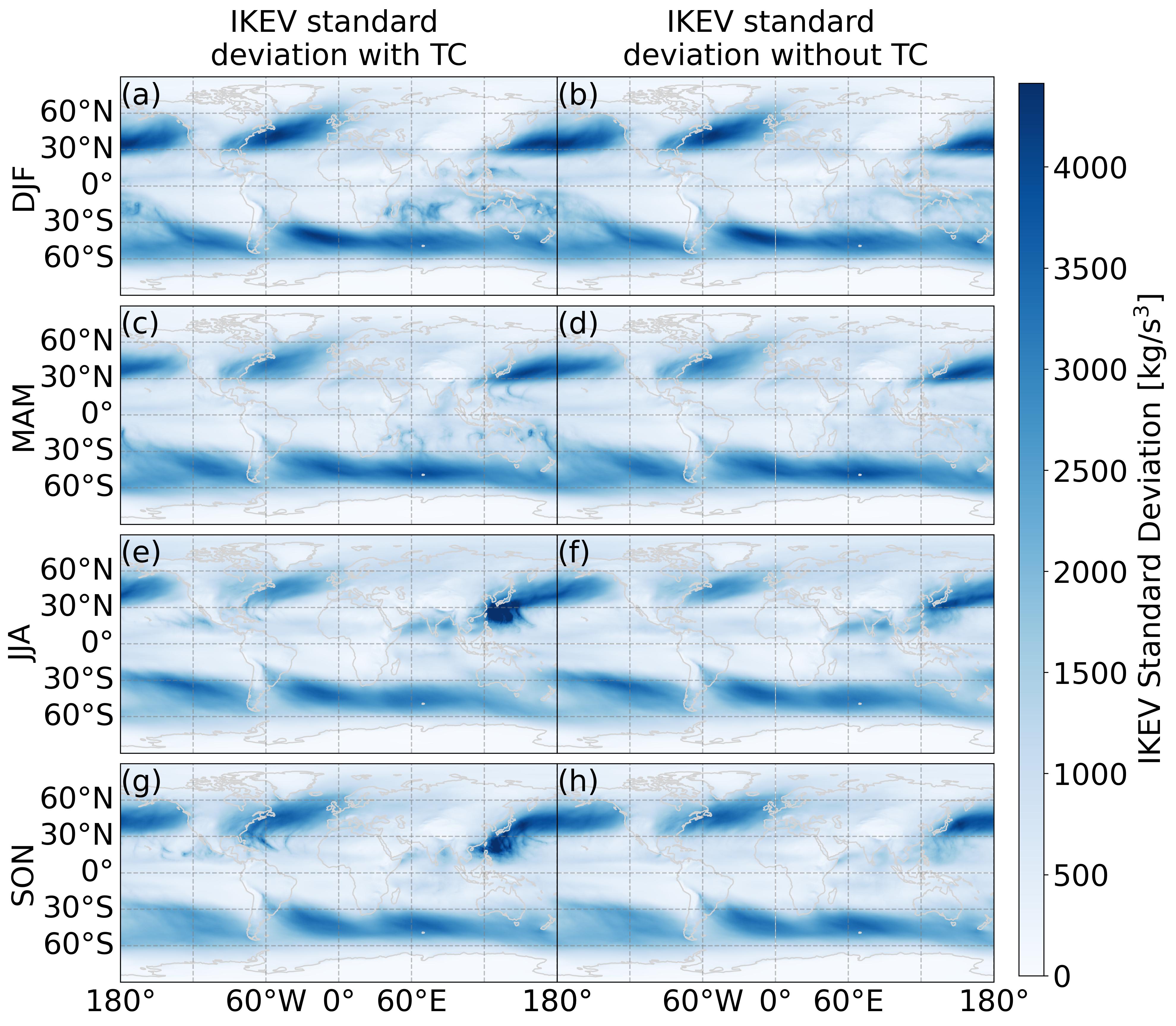}
    \caption{The standard deviation of IKEV before (left) and after (right) removing tropical cyclones (TC). Before the removal, there is strong IKEV in the southwest Pacific and Atlantic in JJA and SON. After explicitly removing TC, the regions of high IKEV standard deviation are in elongated shapes, similar to that of the AR frequency map in figure 1.}
    \label{fig:TC_removal}
\end{figure}

In addition to modifying the AR Laplacian threshold, we also introduce a constraint to exclude TCs and similar fast-moving moist systems that have strong IVT, IKEV, and IVTE. Some AR detection algorithms include additional geometric constraints on the aspect ratio and size of the AR \cite{rutz2019atmospheric,ralph2019artmip}. which can help distinguish between tropical cyclones and ARs. To eliminate the influence of TCs, we require that an AR must be located at least $4^\circ$ away from the center of any recorded tropical cyclone and similar systems. The tropical cyclone data are obtained from the International Best Track Archive for Climate Stewardship (IBTrACS) project \cite{knapp2010international,gahtan2024ibtracs}. We select a $4^\circ$ threshold to exclude the moist core region of TCs, while still allowing moist filaments extending from TCs to be classified as ARs. To demonstrate the difference, figure~\ref{fig:TC_removal} shows the standard deviation of IKEV before (left) and after (right) removing TCs. The difference is the most significant in the TC active regions during its active season (southwest Pacific in JJA and SON, southwest Atlantic in SON, and tropical Indian Ocean and Pacific Ocean in DJF and MAM). After the removal, the IKEV standard deviations in all seasons become a similar shape as the frequent-AR regions (see figure 1), which indicates that the AR is captured without being affected by the tropical cyclones.

\begin{figure}
    \centering
    \includegraphics[width=1\linewidth]{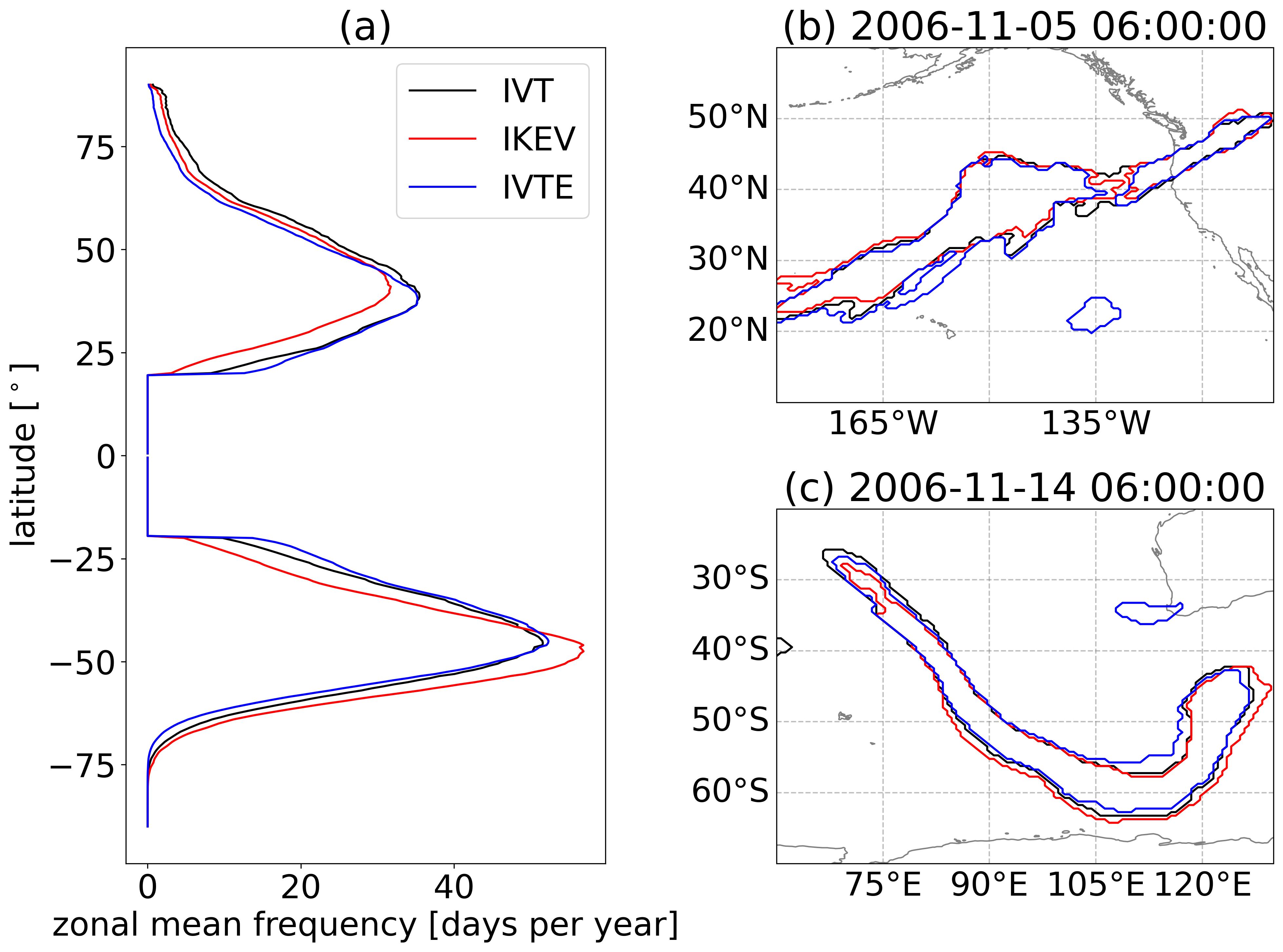}
    \caption{Comparison of AR detections using different AR variables. Panel (a) shows the zonal-mean AR frequency from 2010 to 2019 based on different variables: black curves for IVT, red for IKEV, and blue for IVTE. The tropical region within ±20° latitude is excluded from AR detection. Panels (b) and (c) compare two individual cases; inter-algorithm differences for these cases are discussed in Figure 4 of \citeA{lora2020consensus}. Differences among variables are comparable to those among ARTMIP members.}
    \label{fig:IVs_freq_zonal_mean}
\end{figure}

\begin{figure}
    \centering
    \includegraphics[width=1\linewidth]{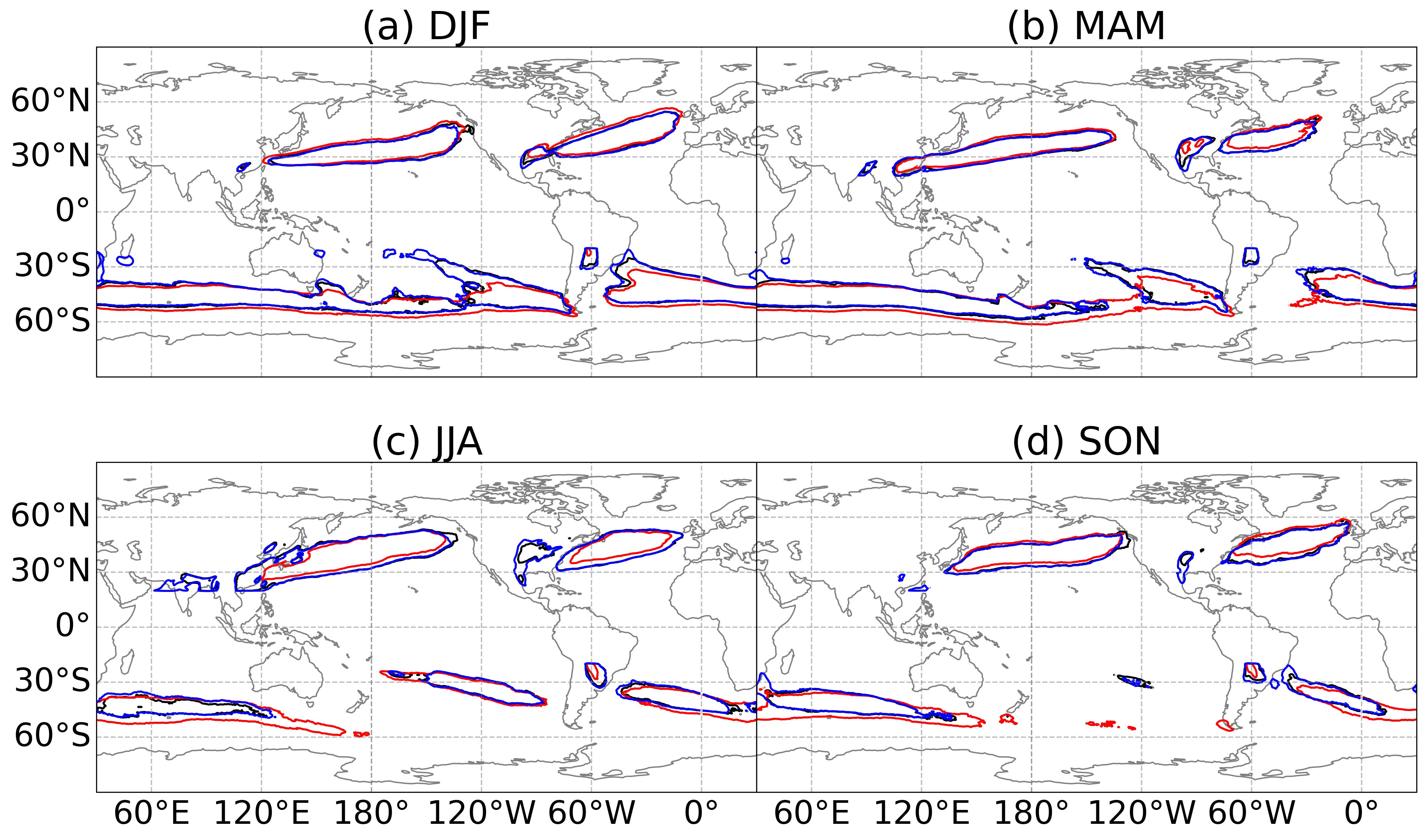}
    \caption{Comparison of seasonal AR frequency using different AR variables. The four panels shows the contours where AR frequency is $15\%$ of days of the season. The black curves are for IVT, reds for IKEV, and blues for IVTE. AR frequencies from IVT and IVTE are very similar in all seasons, whereas the AR frequencies from IKEV capture more features in high latitude (e.g., the Southern Indian Ocean in all seasons), and fewer features from moisture (e.g., the munsoon season in southern and eastern Asia.)}
    \label{fig:IVs_freq_seasonal}
\end{figure}

Figure~\ref{fig:IVs_freq_zonal_mean} compares AR detections using different AR varibles. Panel (a) shows the zonal-mean AR frequency from 2010 to 2019 based on different variables: black curves for IVT, red for IKEV, and blue for IVTE. The tropical region within ±20° latitude is excluded from AR detection. The results of IVT and IVTE are extremely similar because they consider $q$, $u$, and $v$ with equal power. The frequency of IKEV is stronger in the southern hemisphere but weaker in the northern hemisphere because of the strong jet in the southern hemisphere. Panels (b) and (c) compare two individual cases; inter-algorithm differences for these cases are discussed in Figure 4 of \citeA{lora2020consensus}. Differences among variables are comparable to those among ARTMIP members. Figure~\ref{fig:IVs_freq_seasonal} shows the seasonal frequencies using the same AR-detection algorithm and different AR variables (black curves for IVT, red for IKEV, and blue for IVTE). Similarly, IVT/IVTE is more sensitive to mositure features and IKEV is more sensitive to wind features.

\noindent\textbf{Text S2. The Choice of Regions for Global AR Analysis}

\begin{figure}
    \centering
    \includegraphics[width=1\linewidth]{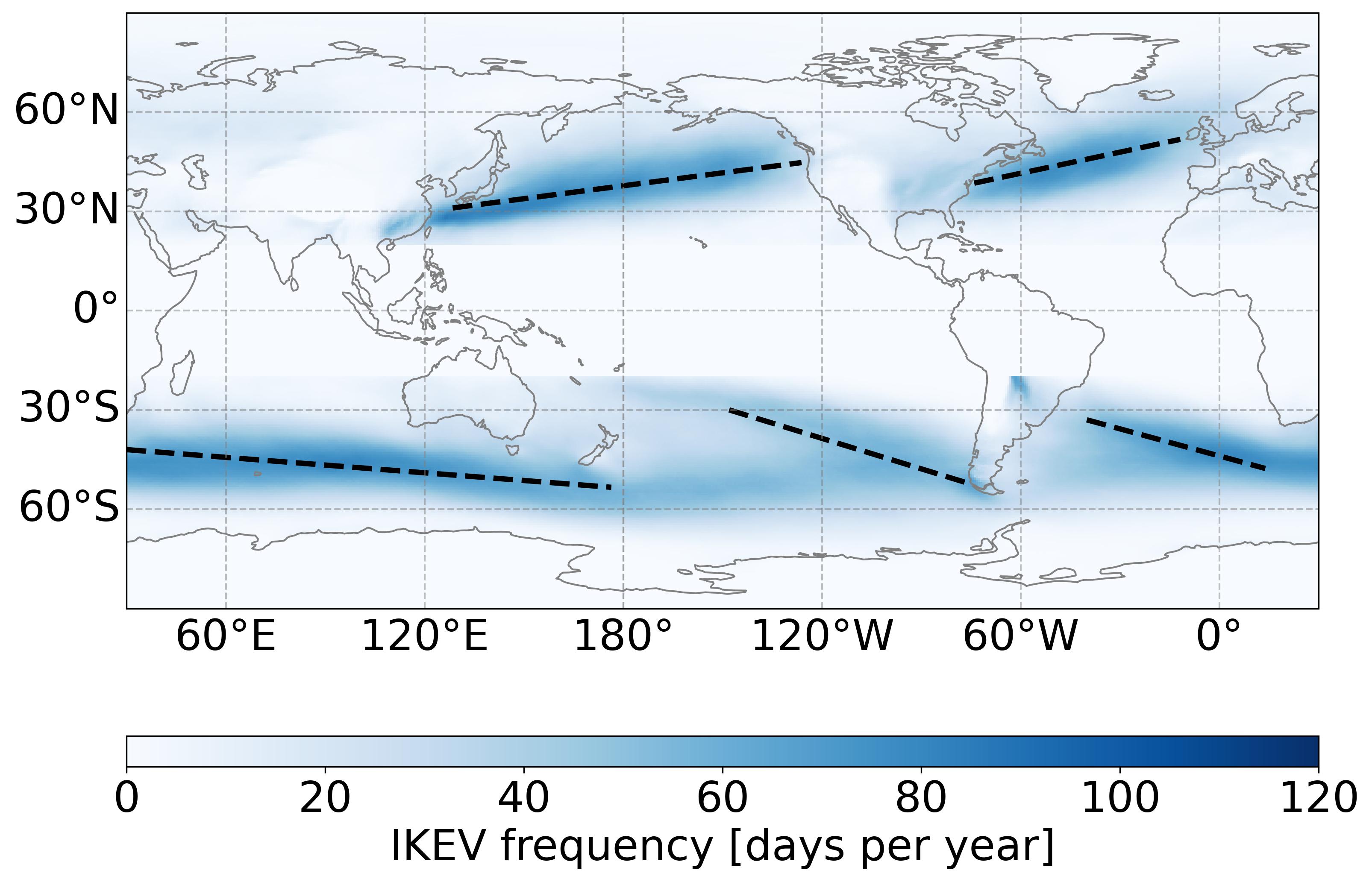}
    \caption{The domain centers used for the global AR budget analysis. The blue color is the annual mean AR frequency from our detection algorithm based on IKEV. The black dashed lines show the domain centers of the composites used in the global budget analysis.}
    \label{fig:comp_paths}
\end{figure}

To perform the AR budget analysis across different regions, we select five regions of frequent AR occurrence, which are also the consensus regions among AR detection algorithms \cite{lora2020consensus}. In each region, we conduct 50 sets of composite analyses by doing linear regressions in a rectangular box (For the budget analysis, the box is $60^\circ$ wide in longitude and $30^\circ$ wide in latitude. For the  Hovmöller analysis, the box is $150^\circ$ wide in longitude and $30^\circ$ wide in latitude.) along the straight line in the region. Figure~\ref{fig:comp_paths} shows the straight lines used as representative AR locations in this study.

Given a representative AR location, the detailed procedure for calculating \textit{one} set of AR composites is described as follows,

\begin{enumerate}
    \item Center the composite domain at the given representative AR location. 
    \item To obtain an AR index variable, we calculate the time series of the mean IKEV (or IVTE) within a $1^\circ\cross 1^\circ$ box (index box) at the domain center. 
    \item Collect the terms of interest, which include the AR variable (KEV or VTE) and the tendency terms from their governing equations (6)--(7). All terms are vertically integrated as we focus on the horizontal variations.
    \item (For the Hovmöller diagram calculation only) Shift each term of interest by the desired lead or lag. Discard boundary days with missing data.
    \item Filter out frames without AR events by removing time frames where no AR is present within the index box within $\pm1$ day.
    \item Perform linear regressions on each term of interest against the AR index variable. Multiply the slope of the regression by the standard deviation of the AR index variable to obtain the dimensionally correct composite of AR tendency terms. Note that the effect of the mean field is removed by not considering the intercept term of the linear regression.
    
    \item Additionally, perform two-tailed Student's t-tests on each grid point, following the approach described in \citeA{ong2024vapor}. Regressions are considered statistically significant at a 95\% confidence level. Statistically insignificant data is set to zero in our budget analysis.
\end{enumerate}

\noindent\textbf{Text S3. Additional Plots for the Budget Analysis}

\begin{figure}
    \centering
    \includegraphics[width=1\linewidth]{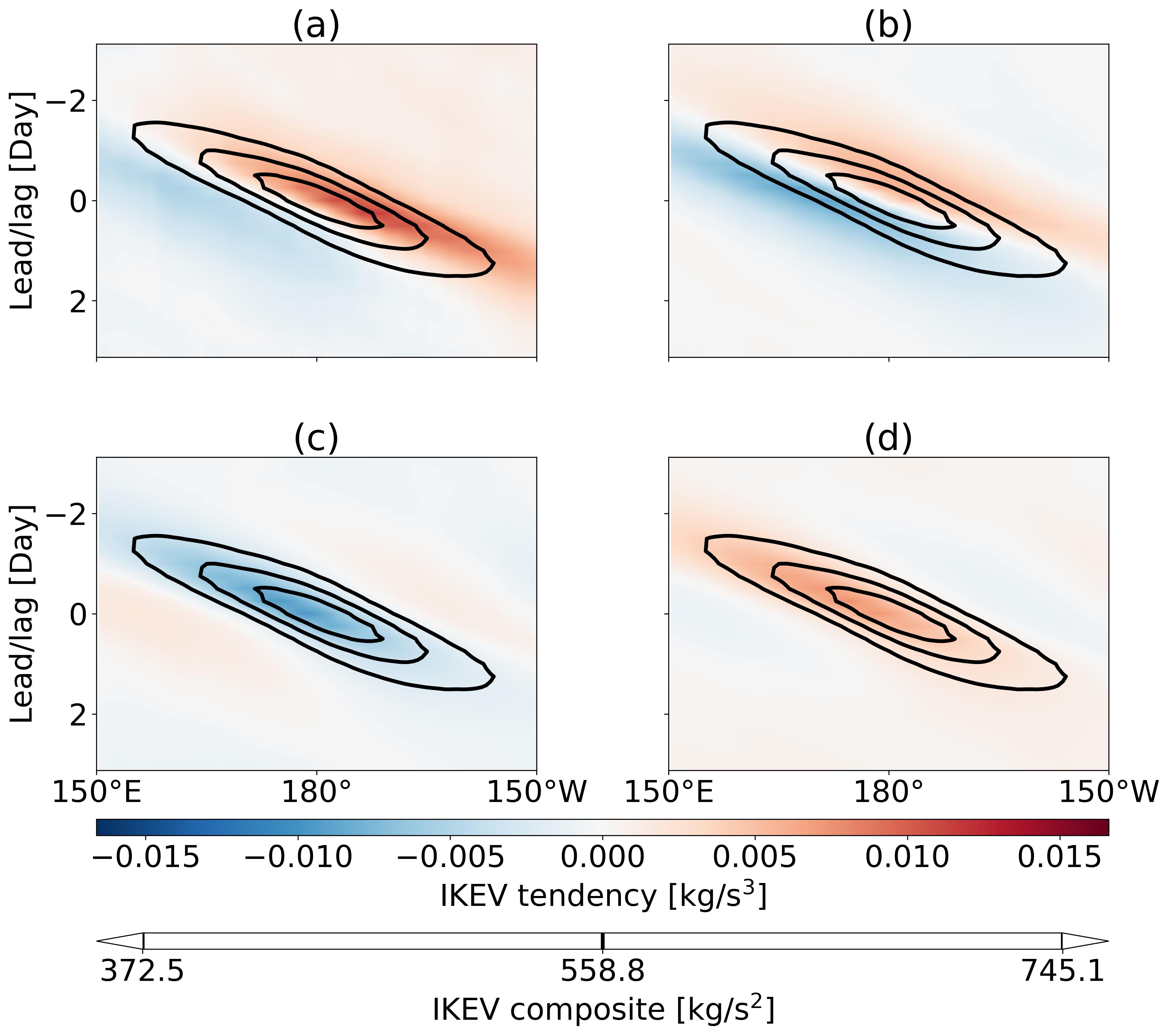}
    \caption{Decomposition of the flux terms in the Hovmöller diagram (Figure 3, the latitudinal-averaged composites of convergence in the North Pacific). The black contours show the mean IKEV composite (The levels are 40\%, 60\%, and 80\% of the maximum value). The colored fields are the regressed IKEV tendency variable. Panel (a) is for horizontal convergence of the KE flux; panel (b) is for horizontal convergence of the vapor flux; panel (c) shows the vertical convergence of the KE flux; panel (d) shows the vertical convergence of the vapor flux.}
    \label{fig:HVM_detailed}
\end{figure}

Figure~\ref{fig:HVM_detailed} provides a detailed decomposition of the flux terms shown in Hovmöller diagram in addition to Figure 2.

\begin{figure}
    \centering
    \includegraphics[width=1\linewidth]{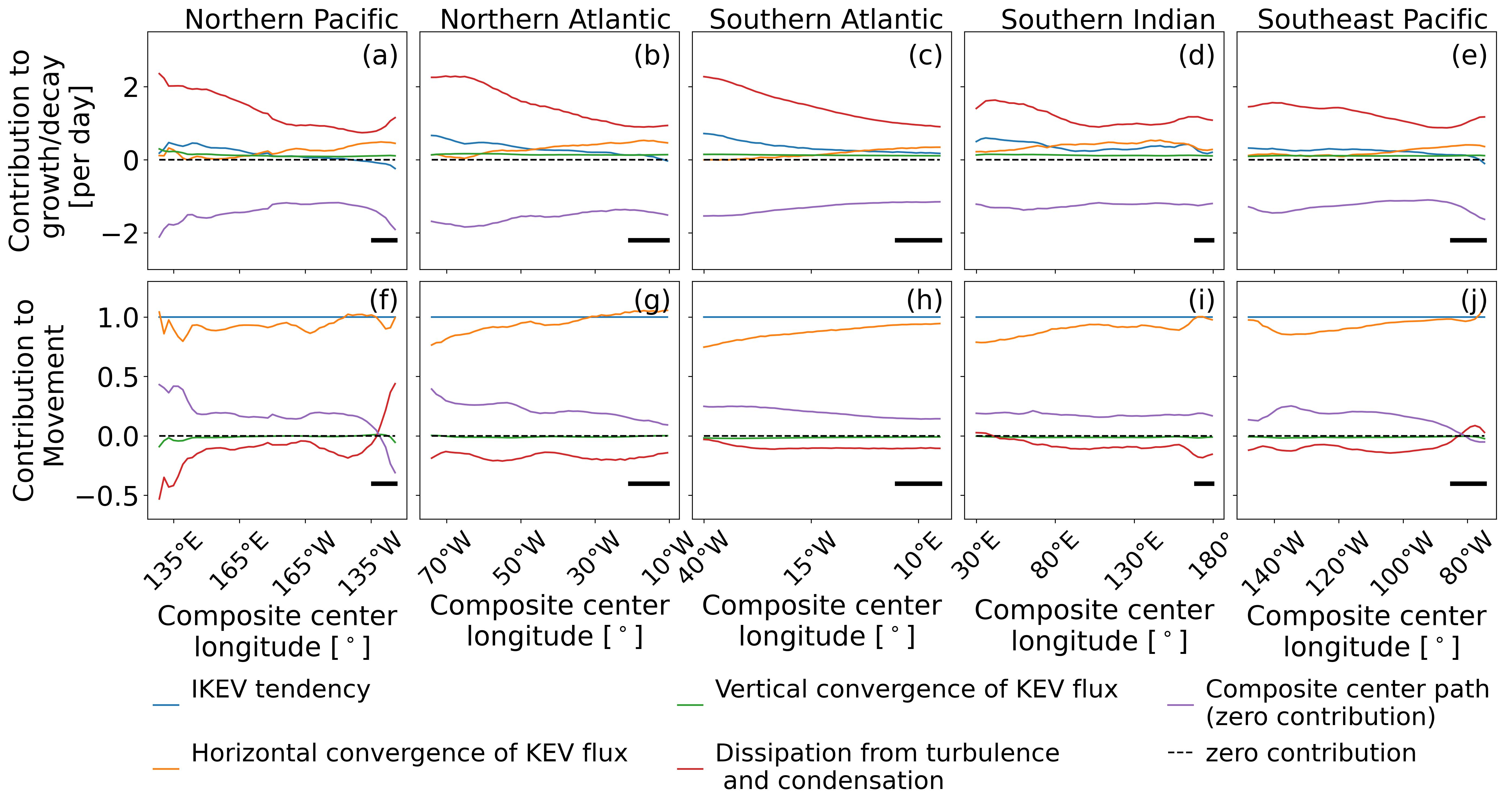}
    \caption{The panel version of the global IKEV budget analysis on the AR evolution. Each column represents different basins. The top panels are for the contribution to the growth/decay. The bottom panels are for the contribution to the movement. The horizontal black bar in the bottom right of each panel stands for $10^\circ$ longitude. The colors correspond to the terms indicated in the legend.}
    \label{fig:3_panelized}
\end{figure}

Figure~\ref{fig:3_panelized} presents a multi-paneled version of Figure~3, which shows the global IKEV budget analysis on the AR evolution. As basin widths differ, horizontal black bars are included in each panel to denote $10^\circ$ of longitude.

\begin{figure}
    \centering
    \includegraphics[width=1\linewidth]{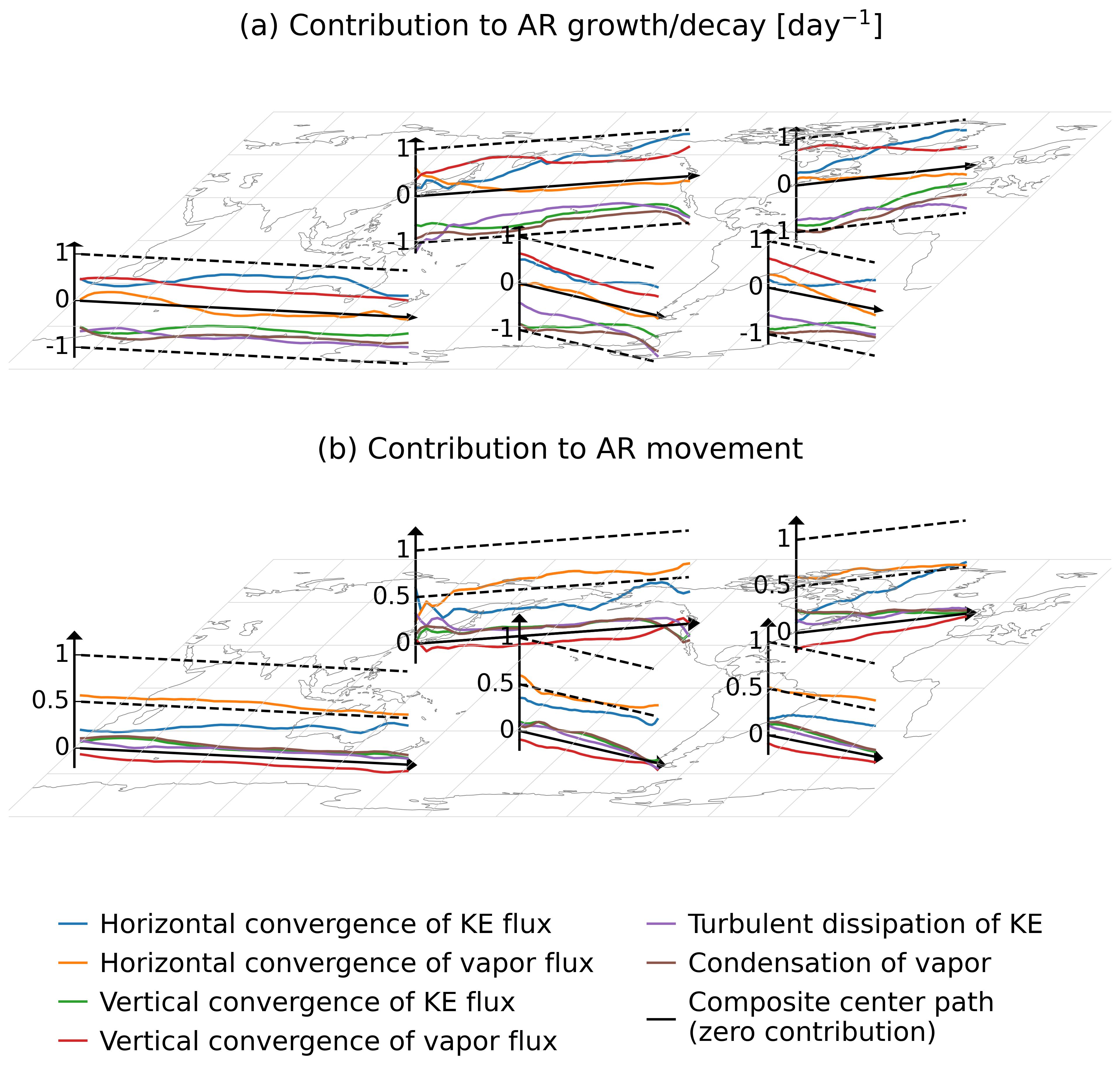}
    \caption{Decomposition of the flux and dissipation the tendency terms in figure 3 (the global IKEV budget analysis on the AR evolution). Panel (a) shows the contribution to the growth/decay. Panel (b) shows the contribution to the movement. Each vertical panel shows the growth/decay contribution from the dominant terms in each basin. The horizontal arrows indicate the center of each composite. The colors correspond to the terms indicated in the legend.}
    \label{fig:Scan_detailed}
\end{figure}

Figure~\ref{fig:Scan_detailed} provides a decomposition of the flux and dissipation terms shown in Figure 3.

\begin{figure}
    \centering
    \includegraphics[width=1\linewidth]{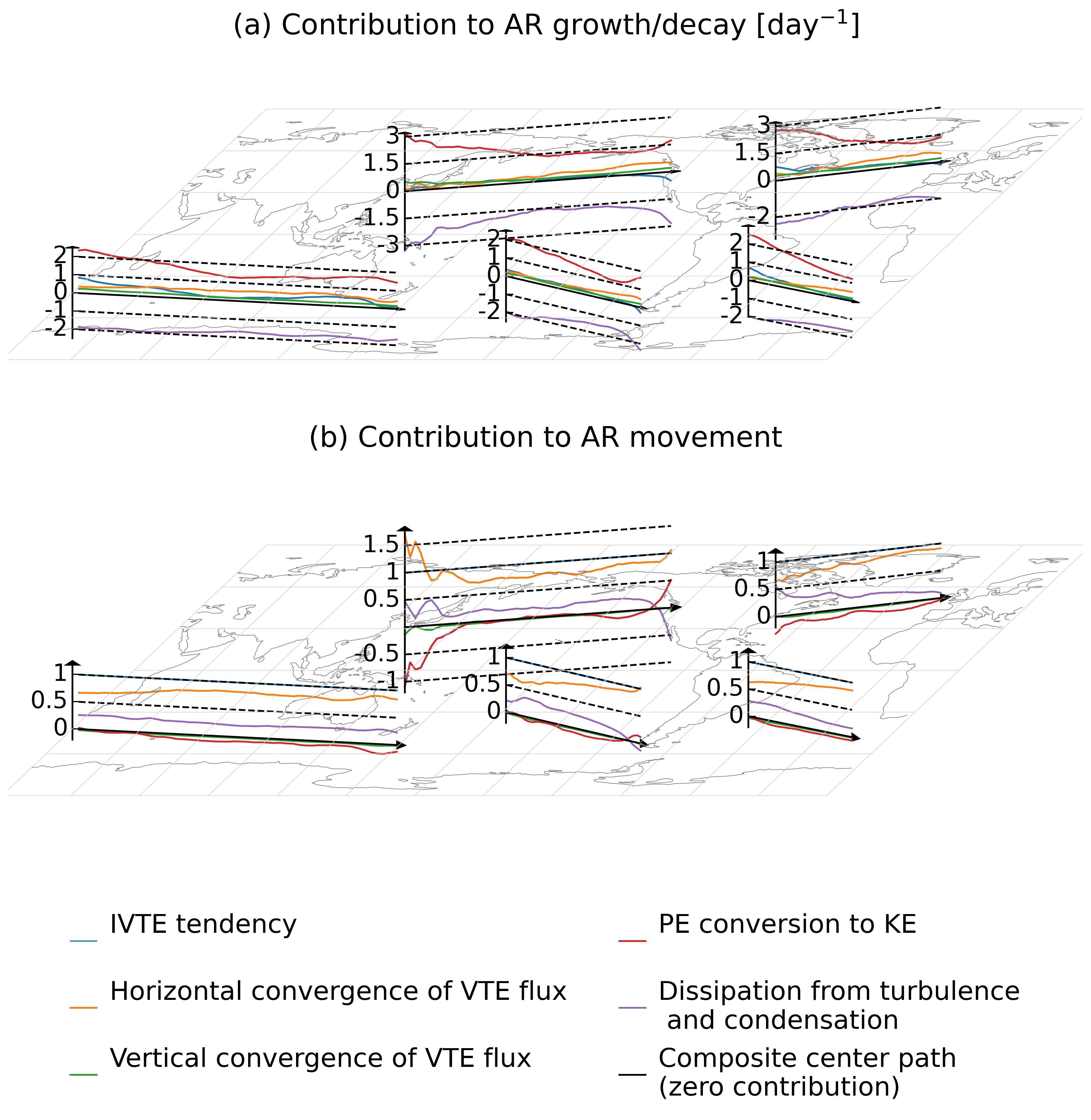}
    \caption{The global IVTE budget analysis on the AR evolution. Panel (a) shows the contribution to the growth/decay. Panel (b) shows the contribution to the movement. Each vertical panel shows the growth/decay contribution from the dominant terms in each basin. The horizontal arrows indicate the center of each composite. The colors correspond to the terms indicated in the legend.}
    \label{fig:Scan_IVTE}
\end{figure}

Figure~\ref{fig:Scan_IVTE} is the budget analysis with IVTE as the AR variable. All findings obtained from Figure 3 still hold when we use IVTE instead of IKEV.

\begin{figure}
    \centering
    \includegraphics[width=1\linewidth]{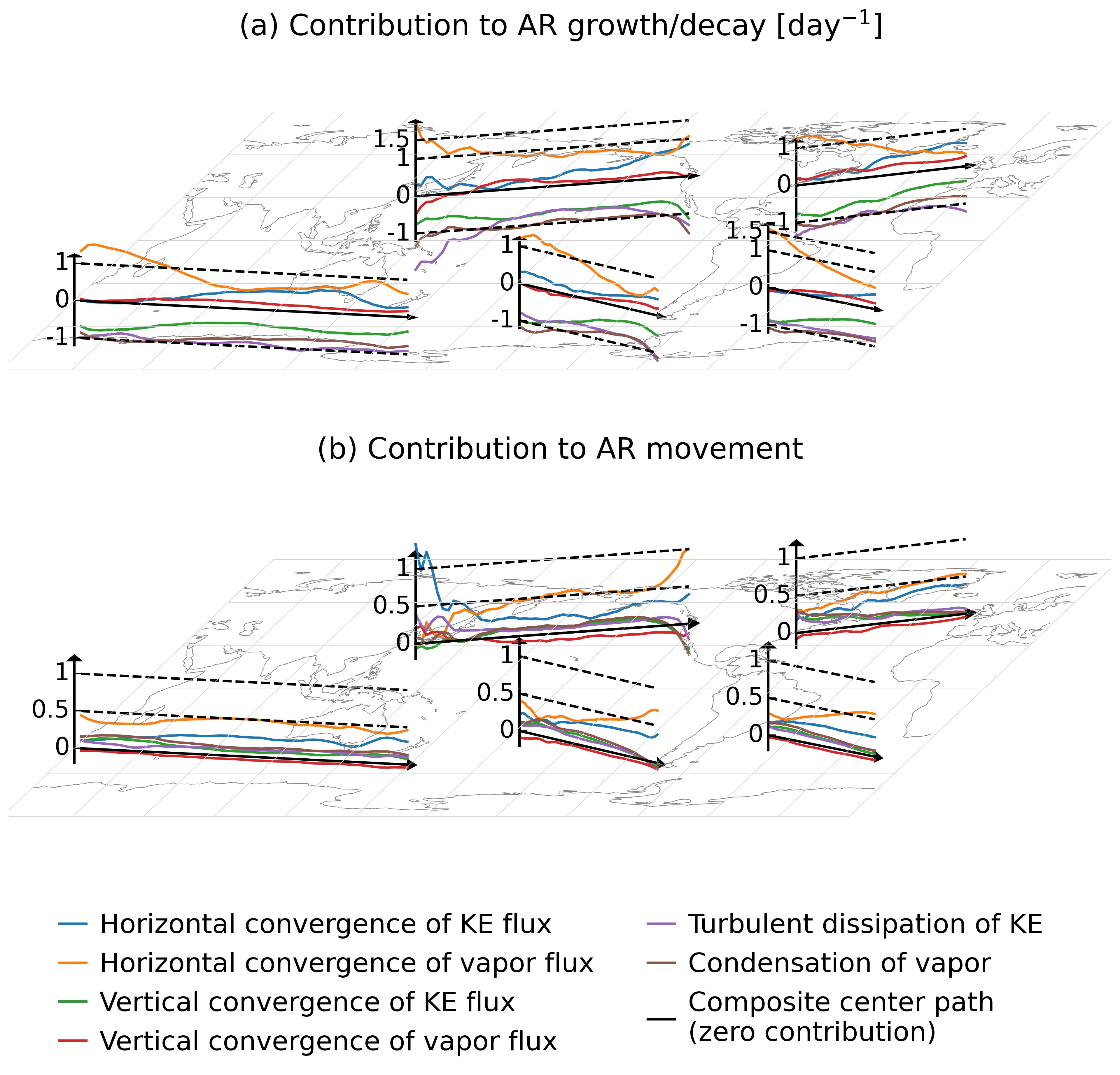}
    \caption{Decomposition of the flux and dissipation the tendency terms in figure~\ref{fig:Scan_IVTE} (the global IVTE budget analysis on the AR evolution). Panel (a) shows the contribution to the growth/decay. Panel (b) shows the contribution to the movement. Each vertical panel shows the growth/decay contribution from the dominant terms in each basin. The horizontal arrows indicate the center of each composite. The colors correspond to the terms indicated in the legend.}
    \label{fig:Scan_IVTE_detailed}
\end{figure}

\begin{figure}[h!]
        \centering
        \includegraphics[width=0.9\linewidth]{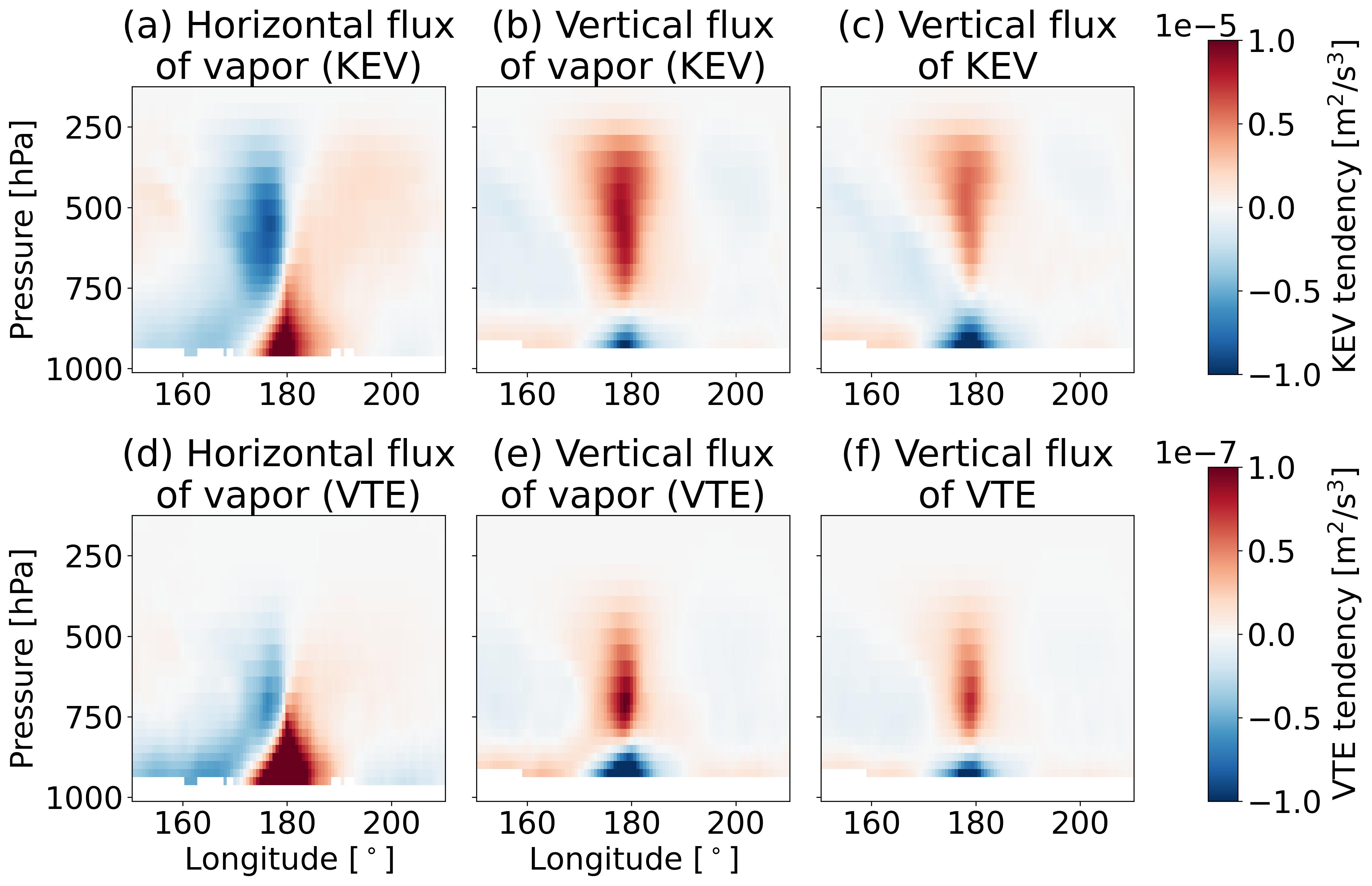}
        
        \caption{The VKE tendency composites in a vertical slice of AR. Panel (a) shows the horizontal flux of vapor in the KEV tendency, $-K\nabp\cdot(q\textbf{u})$. Panel (b) shows the vertical flux of vapor in the KEV tendency, $-K\cdot\pdv{}{p}\,(q\omega)$. Panel (c) shows the vertical flux in the KEV tendency, $-\pdv{}{p}\,(\mathrm{KEV}\cdot\omega)$. Panel (d) shows the horizontal flux of vapor in the VTE tendency, $-2K\nabp\cdot{}{p}\,(q\textbf{u})$. Panel (e) shows the vertical flux of vapor in the VTE tendency, $-2K\cdot\pdv{}{p}\,(q\omega)$. Panel (f) shows the vertical flux in the VTE tendency, $-\pdv{}{p}\,(\mathrm{VTE}\cdot\omega)$.}
        \label{fig:KEV_VTE_ADV_xz}
    \end{figure}

Figure~\ref{fig:Scan_IVTE_detailed} provides a decomposition of the flux and dissipation terms shown in Figure~\ref{fig:Scan_IVTE}. Note that while most patterns in Figure~\ref{fig:Scan_IVTE_detailed} are consistent with Figure~\ref{fig:Scan_detailed}, the contributions from the convergence of vapor flux are different. The vertical vapor flux is stronger than the horizontal vapor flux in the KEV analysis, whereas the former is weaker than the latter in the VTE analysis. The difference arises because these two terms have an opposite-sign vertical structure (see figure~\ref{fig:KEV_VTE_ADV_xz}, panels (a), (b), (d), and (e)). The VTE depends more on the low-level moisture. And the dependence changes the relative magnitude of the low-level and mid-troposphere profile.

Figure S10 also provides an example of the vertical distribution of the vertical convergence of the VKE flux (panel (c) for KEV and panel (f) for VTE). The vetical convergence of the VKE flux redistributes VKE from the lower troposphere to the higher troposphere.

\newpage

\noindent\textbf{Text S4. The Effect of Baroclinic Instability and Extratropical Cyclones}
 
ARs are closely linked to baroclinic instability\cite{zhu1998proposed,payne2020responses,lee2021dynamics}. In this section, we compare regions of frequent AR occurrence with areas of strong baroclinic instability and compare their regional strength.

We quantify regions of strong baroclinic instability using the Eady Growth Rate (EGR), which represents the growth rate of the most unstable baroclinic mode in an $f$-plane quasi-geostrophic flow with uniform stratification and velocity shear (see Section~9.5 of \citeA{vallis2017atmospheric} for a detailed derivation). Although EGR is derived for an idealized system, it is widely used as an indicator of the strength of baroclinic instability in the atmosphere \cite{simmonds2009biases,simmonds2021trends}. In this study, we estimate the EGR following the method described by \citeA{simmonds2021trends},
\begin{eqnarray}
    \mathrm{EGR} &\equiv& 0.3098\frac{|f|}{N}\Big|\pdv{\bf u}{z}\Big|\label{eq:EGR_def}\\
    N&\equiv&\frac{g}{\theta_v}\pdv{\theta_v}{z}\\
    \theta_v&\equiv&T(1+\epsilon q)\Big(\frac{p}{p_{\mathrm{ref}}}\Big)^{-R/C_p}
\end{eqnarray}
Where $f$ is the Coriolis parameter; $N$ is the Brunt–Väisälä frequency; ${\bf u}$ is the horizontal wind vector; $z$ is the vertical coordinate; $\theta_v$ is the virtual potential temperature; $\epsilon=0.608$ for water vapor in the air.

\begin{figure}
    \centering
    \includegraphics[width=1\linewidth]{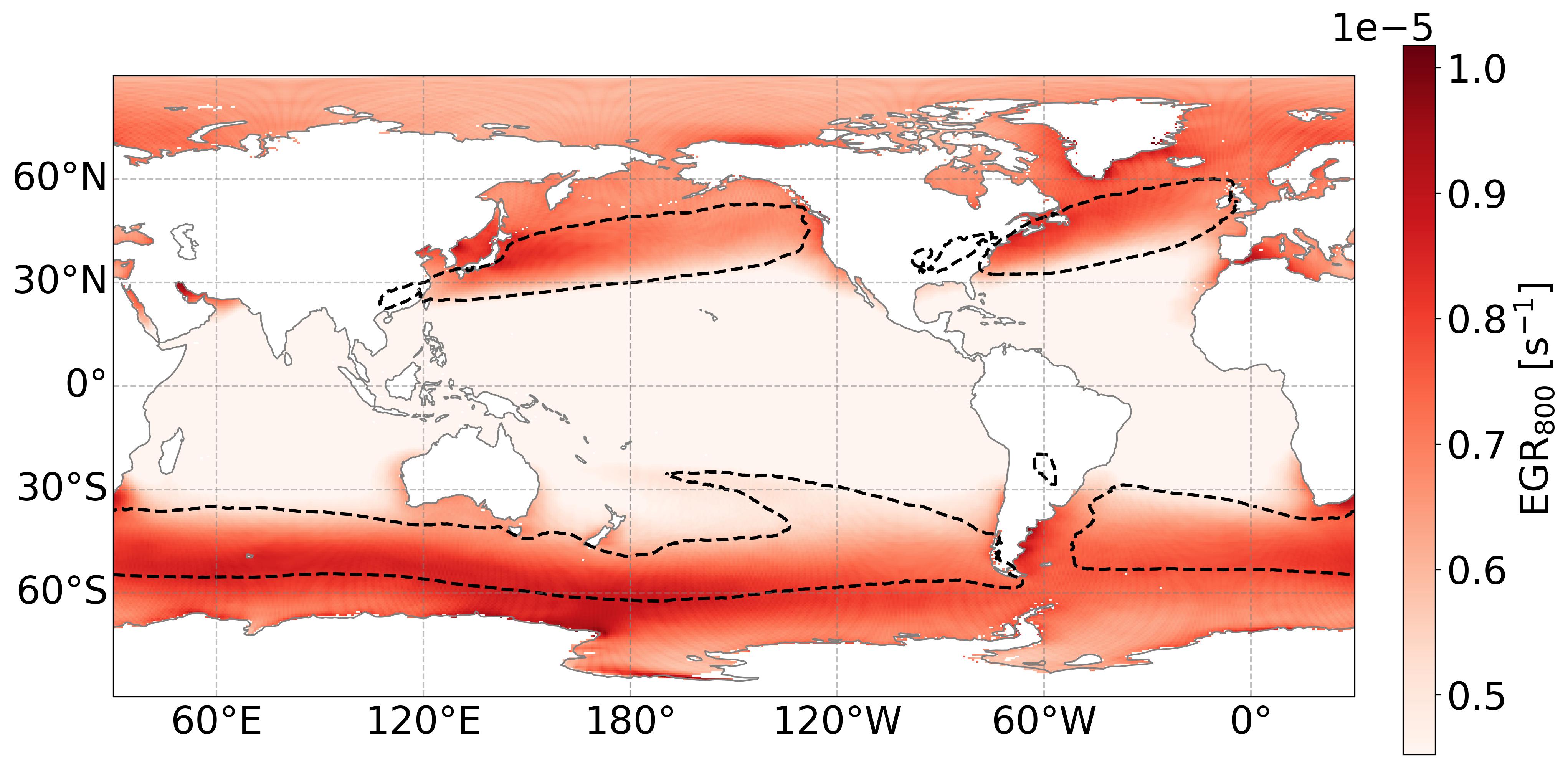}
    \caption{Comparing frequent-AR regions (black contours, measured by IKEV) with the EGR above the ocean at 800~hPa (colored). See text for details. 
    }
    \label{fig:EGR}
\end{figure}

\begin{figure}
    \centering
    \includegraphics[width=1\linewidth]{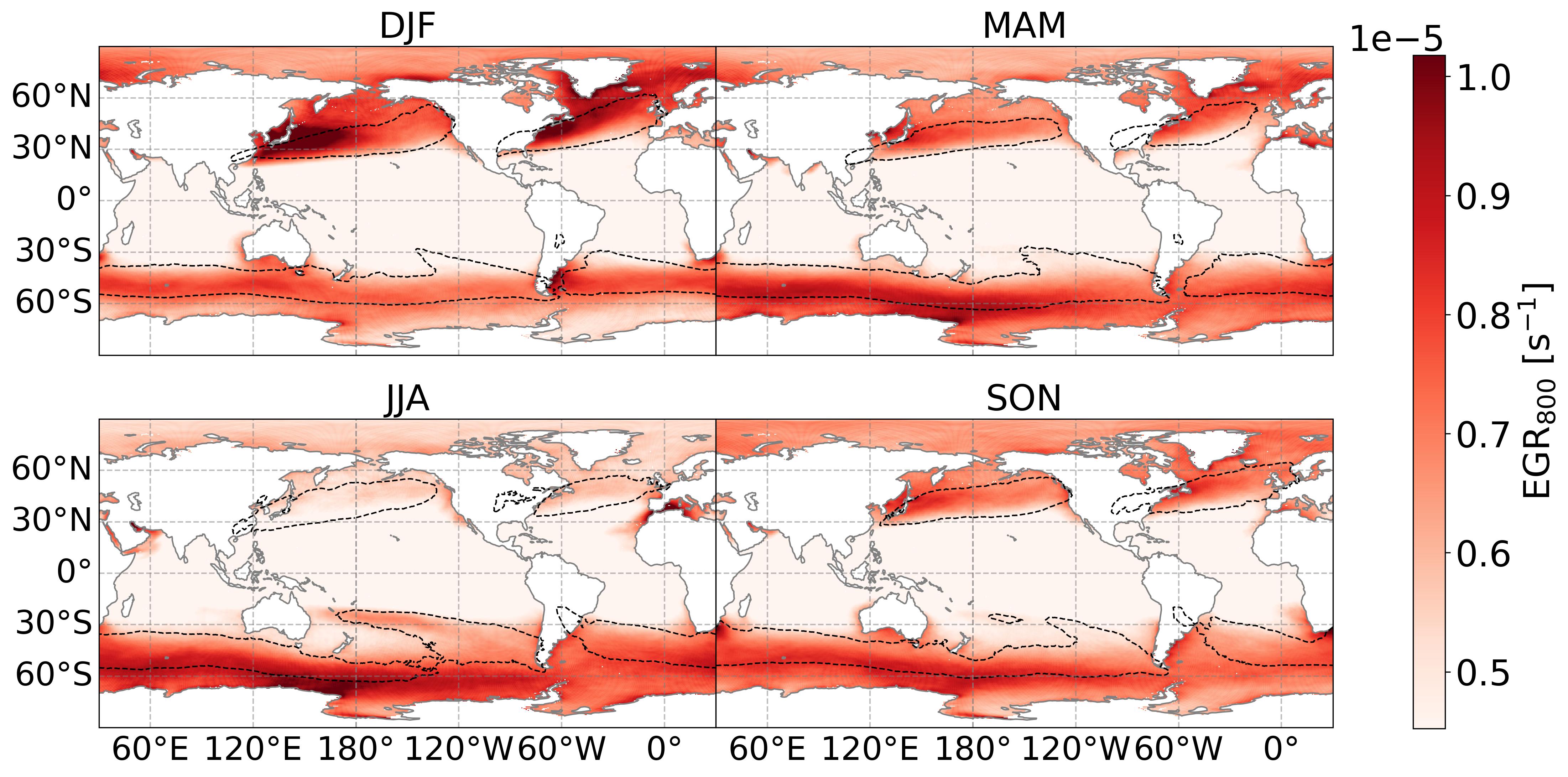}
    \caption{Comparing \textit{seasonal} frequent-AR regions (black contours, measured by IKEV) with the EGR above the ocean at 800~hPa (colored). See text for details. 
    }
    \label{fig:EGR_seasonal}
\end{figure}

Figure~\ref{fig:EGR} compares frequent AR regions (black contours, measured by IKEV) with the EGR distribution at 800~hPa (color map) above the ocean.
The $\mathrm{EGR}_{800}$ is stronger in the west of the basin in adjacent to a major continent (e.g., northwest Pacific, northwest Atlantic, southwest Atlantic, and southern Africa). The west-to-east trend indicates that the atmosphere above the western ocean is more baroclinic unstable than above the eastern ocean, which is consistent with the stronger PE-to-KE conversion in the west basins as shown in figure 3. We note that the $\mathrm{EGR}_{800}$ in the South Pacific is not big. However, this branch is subject to the strong seasonal cycle of the southern hemisphere subtropical jet \cite{gillett2021tropical}. In figure~\ref{fig:EGR_seasonal}, we present the seasonal $\mathrm{EGR}_{800}$ (red coloring) and the frequent AR region (black contours,also see figure~\ref{fig:IVs_freq_seasonal}). In JJA, when there is a strong subtropical jet \cite{gillett2021tropical} and more frequent AR activity in the southern Pacific, there is a branch of strong $\mathrm{EGR}_{800}$ in the region. The branch exhibits a stronger $\mathrm{EGR}_{800}$ at the west of the basin compared with the east. Hence, the spatial variation of $\mathrm{EGR}_{800}$ is consistent with the west-east variation of PE-to-KE conversion.

%%% End of body of article:
%%%%%%%%%%%%%%%%%%%%%%%%%%%%%%%%%%%%%%%%%%%%%%%%%%%%%%%%%%%%%%%%
%
% Optional Notation section goes here
%
% Notation -- End each entry with a period.
% \begin{notation}
% Term & definition.\\
% Second term & second definition.\\
% \end{notation}
%%%%%%%%%%%%%%%%%%%%%%%%%%%%%%%%%%%%%%%%%%%%%%%%%%%%%%%%%%%%%%%%

%% ------------------------------------------------------------------------ %%
%%  REFERENCE LIST AND TEXT CITATIONS

%%%%%%%%%%%%%%%%%%%%%%%%%%%%%%%%%%%%%%%%%%%%%%%
% 
%
% \bibliography{AR} % do not specify file extension

\end{document}